\documentclass[fleqn,usenatbib]{mnras}

\usepackage{newtxtext,newtxmath}

\usepackage[T1]{fontenc}
\usepackage{ae,aecompl}

\usepackage{graphicx}	
\usepackage{amsmath}	

\newcommand{\km}{\mathrm{km}}
\newcommand{\pc}{\mathrm{pc}}

\newcommand{\s}{\mathrm{s}}

\newcommand{\g}{\mbox{g}}
\newcommand{\cm}{\mbox{cm}}
\newcommand{\msol}{\mbox{$\mathrm{M}_{\sun}$}}
\newcommand{\tff}{t_\mathrm{ff}}

\newcommand{\sigvtrue}{\sigma_{v,\mathrm{3D}}}

\newcommand{\sigKD}[1]{\sigma_{\mathrm{c},{#1}}}

\newcommand{\siggradKD}[1]{\sigma_{(\mathrm{c}-\mathrm{grad),{#1}}}}
\newcommand{\siggradKDgen}{\sigma_{(\mathrm{c}-\mathrm{grad)}}}

\newcommand{\sigtwoKD}[1]{\sigma_{\mathrm{i},{#1}}}
\newcommand{\sigtwoKDgen}{\sigma_{\mathrm{i}}}

\newcommand{\sigpKD}[1]{\sigma_{\mathrm{p},{#1}}}

\newcommand{\sigpgradKD}[1]{\sigma_{\mathrm{(p}-\mathrm{grad)},{#1}}}
\newcommand{\sigpgradKDgen}{\sigma_{\mathrm{(p}-\mathrm{grad)}}}

\usepackage{scalerel,tikz}
\usetikzlibrary{svg.path}
\definecolor{orcidlogocol}{HTML}{A6CE39}
\tikzset{orcidlogo/.pic={
 \fill[orcidlogocol] svg{M256,128c0,70.7-57.3,128-128,128C57.3,256,0,198.7,0,128C0,57.3,57.3,0,128,0C198.7,0,256,57.3,256,128z};
 \fill[white] svg{M86.3,186.2H70.9V79.1h15.4v48.4V186.2z}
 svg{M108.9,79.1h41.6c39.6,0,57,28.3,57,53.6c0,27.5-21.5,53.6-56.8,53.6h-41.8V79.1z M124.3,172.4h24.5c34.9,0,42.9-26.5,42.9-39.7c0-21.5-13.7-39.7-43.7-39.7h-23.7V172.4z}
 svg{M88.7,56.8c0,5.5-4.5,10.1-10.1,10.1c-5.6,0-10.1-4.6-10.1-10.1c0-5.6,4.5-10.1,10.1-10.1C84.2,46.7,88.7,51.3,88.7,56.8z};
}}
\newcommand\orcidicon[1]{\href{https://orcid.org/#1}{\mbox{\scalerel*{
\begin{tikzpicture}[yscale=-1,transform shape]
\pic{orcidlogo};
\end{tikzpicture}
}{|}}}}

\title[Measuring 3D turbulence in observed clouds]{A new method for measuring the 3D turbulent velocity dispersion of molecular clouds}

\author[Stewart \& Federrath]{
Madeleine Stewart$^{1}$ \&
Christoph Federrath$^{\orcidicon{0000-0002-0706-2306}\,1}$\thanks{E-mail: christoph.federrath@anu.edu.au}
\\
$^{1}$Research School of Astronomy and Astrophysics, Australian National University, Canberra, ACT~2611, Australia
}

\pubyear{2019}

\begin{document}
\label{firstpage}
\pagerange{\pageref{firstpage}--\pageref{lastpage}}
\maketitle

\begin{abstract}
The structure and star formation activity of a molecular cloud are fundamentally linked to its internal turbulence. However, accurately measuring the turbulent velocity dispersion is challenging due to projection effects and observational limitations, such as telescope resolution, particularly for clouds that include non-turbulent motions, such as large-scale rotation. Here we develop a new method to recover the three-dimensional (3D) turbulent velocity dispersion ($\sigvtrue$) from position-position-velocity (PPV) data. We simulate a rotating, turbulent, collapsing molecular cloud and compare its intrinsic $\sigvtrue$ with three different measures of the velocity dispersion accessible in PPV space: 1) the spatial mean of the 2nd-moment map, $\sigtwoKDgen$, 2) the standard deviation of the gradient/rotation-corrected 1st-moment map, $\siggradKDgen$, and 3) a combination of 1) and 2), called the `gradient-corrected parent velocity dispersion', $\sigpgradKDgen=(\sigtwoKDgen^2+\siggradKDgen^2)^{1/2}$. We show that the gradient correction is crucial in order to recover purely turbulent motions of the cloud, independent of the orientation of the cloud with respect to the line of sight (LOS). We find that with a suitable correction factor and appropriate filters applied to the moment maps, all three statistics can be used to recover $\sigvtrue$, with method~3 being the most robust and reliable. We determine the correction factor as a function of the telescope beam size for different levels of cloud rotation, and find that for a beam FWHM $f$ and cloud radius $R$, the 3D turbulent velocity dispersion can best be recovered from the gradient-corrected parent velocity dispersion via $\sigvtrue = \left[(-0.29\pm0.26)\, f/R + 1.93 \pm 0.15\right] \sigpgradKDgen$ for $f/R < 1$, independent of the level of cloud rotation or LOS orientation.
\end{abstract}




\section{Introduction}
Interstellar turbulence is a key ingredient for determining galaxy structure and evolution, and, in particular, for the formation of stars in molecular clouds \citep{MacLowKlessen2004, McKeeOstriker2007, HennebelleFalgarone2012, PadoanEtAl2014}. Accurately measuring the turbulence of molecular clouds is therefore of interest. For example, the star formation rate depends on the turbulent Mach number, which is determined by the 3D velocity dispersion \citep{PadoanNordlund2011,HennebelleChabrier2011,FederrathKlessen2012,FederrathEtAl2016,Federrath2018}. Likewise, models of the initial mass function depend upon the statistics of turbulence, again with the turbulent velocity dispersion as a key parameter \citep{PadoanNordlund2002,HennebelleChabrier2008,HennebelleChabrier2009,Hopkins2012a,Hopkins2013IMF,NamFederrathKrumholz2021}.
    
The 3D velocity dispersion of a turbulent cloud is quantified by
\begin{equation}
\label{eq: 3d velocity dispersion}
\sigvtrue = \left( \sigma_{v_x}^2 + \sigma_{v_y}^2 + \sigma_{v_z}^2 \right) ^{\frac{1}{2}},
\end{equation}
where $v_x$, $v_y$, and $v_z$ are the gas velocities in the Cartesian coordinate directions $x$, $y$, and $z$, respectively. The obvious challenge in observations is that only 1 of the 3 velocity components is directly measurable, i.e., only the line of sight (LOS) velocity is accessible in PPV observations.
    
Quantifying the turbulence of a rotating molecular cloud becomes even more complex and challenging in observations. The rotational velocity contributes to the dispersion, depending on the orientation of the cloud's rotation axis with respect to the LOS direction towards the cloud, and depending on the spatial resolution of the instrument used to observe the cloud. Importantly, large-scale rotation of a cloud is a systematic motion, and is different from turbulent motions that we are actually interested in measuring, and hence, the contribution of rotation must not be included in a measurement of the turbulence. Therefore, the 3D velocity dispersion that includes rotation or large-scale shearing motions overestimates the true turbulent dispersion, and we have to subtract or correct for the contribution from rotation if we want to isolate the intrinsic 3D turbulent velocity dispersion of a cloud. We define the 3D rotation-corrected velocity dispersion as
\begin{equation}
\label{eq 3d velocity dispersion rotation corrected}
\sigvtrue = \left( \sigma_{v_x - v_{\mathrm{rot} \> x}}^2 + \sigma_{v_y - v_{\mathrm{rot} \> y}}^2 + \sigma_{v_{z - v_{\mathrm{rot} \> z}}}^2 \right) ^{\frac{1}{2}},
\end{equation}
where $v_{\mathrm{rot} \> x}$, $v_{\mathrm{rot} \> y}$, and $v_{\mathrm{rot} \> z}$ are the $x$, $y$, and $z$-components of the rotational velocity, respectively. This statistic represents the purely turbulent motions of a cloud, and is the statistic we would like to obtain from PPV data.
    
Neither the 3D velocity dispersion nor the 3D rotation-corrected velocity dispersion can be directly measured from observational data; only velocities along the LOS are measurable via the Doppler shift of emission or absorption lines. Thus, a suitable correction factor needs to be applied, which converts the 1D measured velocity dispersion to the 3D dispersion. Here we aim to calibrate this correction factor. The spatial mean of the 2nd-moment map is the most commonly used statistic to quantify the turbulence of a molecular cloud from observational data \citep{Larson1981,OssenkopfMacLow2002}. To compensate for the lack of information of the velocity components in the plane of the sky, one usually has to assume isotropy to infer the 3D turbulence. Like the 3D velocity dispersion, these measurements may include contributions from the systematic motion of a rotating or shearing cloud, so we also need to correct for the effects of rotation, by determining a suitable correction factor.
    
An alternative method to using the 2nd moment, which corrects for the contributions from systematic motions (e.g., rotation) of a cloud, was presented in \citet{FederrathEtAl2016}. Rotational and shearing motions appear in the 1st-moment map of a cloud as a gradient, unless the LOS is along the rotation axis \citep{MyersBenson1983, GoldsmithArquilla1985}. This gradient can be fit to, and subtracted from, the velocity map of the cloud to isolate the turbulent motions \citep{FederrathEtAl2016,ShardaEtAl2018,ShardaEtAl2019,MenonEtAl2021}. The standard deviation of the resulting velocity map can then be used as a measurement of turbulence. Although this method corrects for contributions from systematic motion to the velocity dispersion, it fails to include contributions to the velocity dispersion along the LOS, and may thus underestimate the 3D turbulence.

\citet{KleinerDickman1984,KleinerDickman1985} and \citet{DickmanKleiner1985} showed that the total velocity dispersion of a cloud can be found by adding the internal velocity dispersion (the spatial mean of the 2nd-moment map) and the centroid velocity dispersion (the standard deviation of the 1st-moment map) in quadrature. However, this method includes contributions from large-scale rotation and shear, i.e., non-turbulent contributions. We therefore modify this statistic to introduce a new method for measuring the 3D turbulence of a cloud that corrects for systematic motion: the gradient-corrected parent velocity dispersion, which is calculated by adding the internal velocity dispersion and the standard deviation of the gradient-corrected 1st-moment map in quadrature.

Our overarching aim is to reconstruct the intrinsic 3D turbulent (and thus rotation-corrected) velocity dispersion of a cloud from the quantifiers of velocity dispersion obtainable from PPV data by comparing the methods outlined above: 1) the spatial mean of the 2nd-moment map, 2) the standard deviation of the rotation-corrected 1st-moment map, and 3) the gradient-corrected parent velocity dispersion as an extension of the parent velocity dispersion introduced in \citet{KleinerDickman1984,KleinerDickman1985} and \citet{DickmanKleiner1985}. Motivated by this, we investigate the turbulence of an idealised simulated rotating molecular cloud. At each time step in the cloud's evolution, we calculate the 3D rotation-corrected velocity dispersion and the three turbulence quantifiers introduced above, as they would be measured along the $x$, $y$, and $z$ axes in synthetic observations. We determine the relationships between these observable statistics and the intrinsic 3D turbulent velocity dispersion. This, in combination with other methods that reconstruct 3D density and velocity statistics from PPV data \citep[e.g.,][]{BruntFederrathPrice2010a,BruntFederrathPrice2010b,BurkhartLazarian2012,BruntFederrath2014,KainulainenFederrathHenning2014}, is crucial for advancing our understanding of interstellar clouds.

The rest of this work is organised as follows. The simulation method and definitions of the relevant statistics are given in Sec.~\ref{sec: methods section}. The intensity (zero-moment) maps, 1st-moment maps, and 2nd-moment maps are discussed in relation to the evolution and rotation of the simulated cloud in Sec.~\ref{sec: moment maps}. The rotation-corrected 1st-moment map and the difference between velocity probability distribution functions (PDFs) before and after the rotation correction are investigated in Sec.~\ref{sec: gradients and first moments y-axis}. The different turbulence quantifiers and the `true' turbulence of the cloud are then examined as a function of time in Sec.~\ref{sec: tevol}. In Sec.~\ref{sec: inferring turbulence} we explore methods to infer the 3D turbulence from the observable turbulence quantifiers introduced above. Specifically we examine the `correction factor' of each turbulence quantifier; the factor which recovers the `true' 3D turbulence of the cloud when multiplied by the respective statistic. To ensure versatility of the correction factors found, the effects of the telescope resolution and the strength of cloud rotation are investigated in Sec.~\ref{sec: effects of resolution} and Sec.~\ref{sec: effects of strength of rotation}, respectively. Finally, our conclusions are summarised in Sec.~\ref{sec: conclusion}.

\section{Methods} \label{sec: methods section}

\subsection{Numerical simulations} \label{sec: simulations}
    
We use a modified version of the adaptive mesh refinement (AMR) \citep{BergerColella1989} code FLASH \citep{FryxellEtAl2000,DubeyEtAl2008} (v4) to solve the three-dimensional, compressible hydrodynamical equations including self-gravity \citep{Ricker2008,WuenschEtAl2018}. Our basic setup consists of a spherical cloud of radius $R=1\,\pc$ and mean density $\rho_0=1.62\times 10^{-20}\,\g\,\cm^{-3}$, which gives a total cloud mass of $1000\,\msol$. The cloud is initialised at the centre of a cubical, Cartesian computational domain of side length $L=2.2\,\pc$. The temperature of the cloud is set to $T=10\,\mathrm{K}$. The surrounding medium is given a density $\rho_0/100$ and a temperature $T\times100$, in order to set up the cloud in pressure equilibrium with its low-density surrounding medium.
    
Similar to the simulations presented in \citet{FederrathBanerjeeClarkKlessen2010}, \citet{FederrathEtAl2014}, \citet{GerrardFederrathKuruwita2019}, \citet{KuruwitaFederrath2019}, and \citet{KuruwitaFederrathHaugboelle2020}, we include rotation along the $z$-axis. Here we use an angular frequency of $\Omega=5.124 \times 10^{-14}\,\s^{-1}$, which corresponds to a rotational parameter of $\Omega\times\tff = 0.85$, with the freefall time $\tff=3\pi/(32 G \rho_0) = 0.52\,\mathrm{Myr}$, or a ratio of rotational to gravitational energy of $E_\mathrm{rot} / E_\mathrm{grav} = 0.19$.
    
Turbulence is included by constructing a Gaussian random vector field for the three velocity components, following the Fourier method described in \citet{FederrathDuvalKlessenSchmidtMacLow2010}. The power spectrum of the initial velocity fluctuations follows a $P_v(k) \propto k^{-2}$ spectrum, in the wavenumber range $2 \leq k/(2\pi/L) \leq 20$, consistent with the observed velocity scaling in molecular clouds \citep{Larson1981,SolomonEtAl1987,OssenkopfMacLow2002,HeyerBrunt2004,RomanDuvalEtAl2011} and the statistics of supersonic turbulence \citep{KritsukEtAl2007,Federrath2013,FederrathEtAl2021}. Here we set the velocity dispersion on scale $L/2$ (i.e., for the entire, cubical computational domain) to $1\,\km\,\s^{-1}$. Although the cloud occupies a spherical volume of radius $R=1\,\mathrm{pc}$ instead of the whole computational box, the initial 3D velocity dispersion inside the cloud is still $1\,\km\,\s^{-1}$, differing by less than $2\%$ from the box-scale velocity dispersion, since most of the dispersion is on large scales covered by the cloud. This value corresponds to a ratio of turbulent-to-rotational energy of $E_\mathrm{turb} / E_\mathrm{rot} = 1.0$. Thus, the cloud's rotational energy is equal to the turbulent energy. This is a relatively extreme case of high rotation, but we choose this in order to study the most difficult case for disentangling turbulent from rotational (systematic) motions. However, in Sec.~\ref{sec: effects of strength of rotation} we study the effect of different levels of rotation and find that the preferred method for recovering the 3D turbulent dispersion from PPV data is largely insensitive to the rotation of the cloud.

For simplicity, star formation \citep[via sink particles; see][]{FederrathBanerjeeClarkKlessen2010} was not included in the simulations, and a uniform grid resolution with a cell size of $0.0086\,\pc$ was chosen. These methods are sufficient to set up a rotating, turbulent cloud for which we can make converged synthetic observations to quantify the correction factors necessary in the conversion from PPV data to intrinsic 3D velocity dispersion statistics.

\subsection{Definition and calculation of velocity dispersions}
Synthetic observations of the simulation data were made along the $x$, $y$, and $z$ axes at different time steps. The simulation data was processed using yt \citep{TurkEtAl2011} to create a uniform 3D grid of cells, each containing the density and the three velocity components of the cloud. In the following definitions, we denote the dimensions of this 3D grid as $(N_x, N_y, N_z) = (256, 256, 256)$, i.e., identical to the intrinsic resolution of the simulation.
    
Before calculating each of the following statistics, the cells within the cloud were first identified according to the gas density in each cell. The computational cells with a density less than 10\% of the original cloud density, i.e., $0.1\rho_0$ (defined in the simulation parameters) were excluded from the analysis. This was done to ensure the following statistics were calculated for cloud matter only, excluding the ambient low-density medium. The complete analysis code including the calculations of moment maps and the velocity dispersions (see below) is available at \url{https://bitbucket.org/Madeleine122333/2ndascscripts}.

\subsubsection{3D rotation-corrected turbulent velocity dispersion \label{sec: true turbulence definition}}
The true turbulence of the cloud is given by the 3D rotation-corrected velocity dispersion. This statistic is not observable, but is the one we aim to recover from PPV data. The calculation of the 3D rotation-corrected velocity dispersion presented here is only applicable for clouds oriented such that the rotation axis is the $z$-axis, but since this is the case for our particular simulation, we chose this specific calculation in favour of a more general derivation. The generalisation to arbitrary rotation axes is straightforward.

The rotational velocity of each cell $i$ within the cloud is first calculated as
\begin{equation}
\label{eq rotational velocity}
\mathbf{v}_{\mathrm{rot} \> i} = \Omega \,
\begin{bmatrix}
-Y\\X
\end{bmatrix} \quad \forall i,
\end{equation}
where $\Omega$ is the angular frequency as defined in the simulation parameters, and $X$ and $Y$ are the $x$ and $y$ components of the displacement vector from the centre of mass to each cell $i$. We then subtract $\mathbf{v}_{\mathrm{rot} \> i}$ from the original $x$ and $y$ velocity components in each cell. Thus, the 3D rotation-corrected velocity dispersion is calculated as
\begin{equation}
\label{eq 3d velocity dispersion rotation corrected 2}
\sigvtrue = \left( \sigma_{v_x - v_{\mathrm{rot} \> x}}^2 + \sigma_{v_y - v_{\mathrm{rot} \> y}}^2 + \sigma_{v_z}^2 \right) ^{\frac{1}{2}},
\end{equation}
where $v_{\mathrm{rot} \> x}$ and $v_{\mathrm{rot} \> y}$ are the $x$ and $y$ components of the rotational velocity, respectively.

\subsubsection{Zero moment}
The zero-moment map represents the column density map of the cloud, which is directly observable using telescopes. It is important in quantifying the structure of interstellar clouds, including shock waves, filaments \citep[e.g.,][]{ArzoumanianEtAl2011,AndreEtAl2014} and the column density distribution \citep[e.g.,][]{SchneiderEtAl2013,KainulainenFederrathHenning2014}. The zero moment of the simulation data is calculated for every pixel $p$ as viewed along a LOS axis $\mathrm{los}=[x,y,z]$ as
\begin{equation}
\label{eq zero moment}
\rho_p = \frac{\sum_{i = 1} ^{N_\mathrm{los}} m_i}{\sum _{i=1} ^{N_\mathrm{los}} V_i} \quad \forall p,
\end{equation}
where $m_i$ and $V_i$ are the mass of gas and volume of cell $i$, respectively, and $N_\mathrm{los}$ is the number of cells along the LOS. This is similar to a zero-moment map obtained in observations, which is an integral over the velocity axis of the PPV data cube, with the difference that Eq.~(\ref{eq zero moment}) calculates the average gas density along the LOS instead of the column density. However, the average gas density is proportional to the column density, and for the purposes of this study, the actual units of the zero-moment map are irrelevant. What matters is that the synthetic moment maps obtained from the simulations represent typical moment maps obtained from PPV observations. In the following we assume optically thin gas, such that opacity effects do not play a role and we can use density-weighted integrals over the velocity distribution to mimic the intensity weighting, i.e., in the optically thin limit, the intensity is directly proportional to the gas mass along the LOS, which is the approximation used in Eq.~(\ref{eq zero moment}).

\subsubsection{Cutoff based on zero moment}
After calculating each moment map, we apply a cutoff whereby every pixel of each map whose corresponding zero moment is less than a specified fraction of the average zero moment is excluded from the analysis. This is done to remove the effects of less dense regions with a low signal-to-noise ratio (SNR). The same procedure is typically applied in observations, i.e., only data above a certain SNR threshold, usually equivalent to a column-density threshold, is retained for further analyses. Here we use a cutoff of $0.1$, i.e., we only study data with a column density $\geq10\%$ of the mean column density. We also vary this threshold and show that a cutoff of $1$ (i.e., retaining only data above the mean column density) yields similar results, which is discussed in Appendix~\ref{sec: appendix cutoff 1}.

\subsubsection{First moment} \label{sec:firstmomdef}
The first moment is the intensity-weighted average velocity along the LOS. The velocity map of a cloud can be used to deduce information concerning the bulk motion of the cloud as well as smaller-scale turbulent motions, which are intrinsically linked to the evolution of the cloud. The first moment is calculated for every pixel $p$ in the plane of the sky by summing over all spectral (velocity) cells (channels) along a LOS axis $\mathrm{los}=[x,y,z]$ as
\begin{equation}
\label{eq: first moment map}
\langle v_\mathrm{los} \rangle_p = \frac{\sum_{i = 1} ^{N_\mathrm{los}} \rho_i \, v_{\mathrm{los}\>i}}{\sum _{i = 1} ^{N_\mathrm{los}} \rho_i} \quad \forall p,
\end{equation}
where $\rho_i$ is the density of cell $i$ and $v_{\mathrm{los}\>i}$ is the velocity in the direction of the LOS of the gas in cell $i$. We note that since the volume of each cell is constant, we can directly use the density in the sums in the numerator and the denominator (otherwise, one would need to replace $\rho_i$ with $m_i$).

We calculate the standard deviation of the 1st-moment map along a LOS, $\sigKD{\mathrm{los}}$, as in \citet{KleinerDickman1985} (index `c' stands for `centroid'),  as the standard deviation of $\langle v_\mathrm{los} \rangle_p$ over all pixels $p$. We do not consider this as a pure quantifier of turbulence as this both neglects the velocity dispersion along the LOS and includes contributions from the rotational velocity of the cloud. The first moment is calculated in order to apply a gradient fit (see below), and the standard deviation of it is determined for comparison to the standard deviation of the gradient-corrected 1st-moment map (discussed in Sec.~\ref{sec:rotcorr1stmom} below).

\subsubsection{Second moment} \label{sec:secondmomdef}
The second moment measures the velocity dispersion along the LOS. It is the basis for the most commonly used statistic to quantify turbulence \citep[e.g.,][]{Larson1981,SolomonEtAl1987,OssenkopfMacLow2002}. In analogy to Eq.~(\ref{eq: first moment map}), the 2nd-moment map is calculated as
\begin{equation}
\label{eq statistical second moment}
\langle v_\mathrm{los}^2 \rangle_p = \frac{\sum_{i = 1} ^{N_\mathrm{los}} \rho_i \, v_{\mathrm{los}\>i} ^2}{\sum _{i = 1} ^{N_\mathrm{los}} \rho_i} \quad \forall p.
\end{equation}

We then calculate the 2nd-moment (dispersion) map as
\begin{equation}
\label{eq second moment}
\sigma_{v_\mathrm{los} \> p} = \left( \langle v_\mathrm{los}^2 \rangle_p - \langle v_\mathrm{los} \rangle ^2 _p \right) ^{\frac{1}{2}} \quad \forall p.
\end{equation}
        
We also calculate the spatial mean (i.e., the average over all pixels $p$) of the 2nd-moment map as in \citet{KleinerDickman1985} (index `i' stands for `internal' velocity dispersion),
\begin{equation}
\label{eq second moment mean}
\sigtwoKD{\mathrm{los}} = \sqrt{\frac{\sum _p \sigma_{v_\mathrm{los} \> p}^2}{N_p}},
\end{equation}
where $N_p$ is the number of pixels in the 2nd-moment map. This method accounts for the variation of the LOS velocities along the LOS, but it does not fully account for LOS velocity variations in the plane of the sky, and may also include contributions from non-turbulent, systematic motions of the cloud, depending on the spatial resolution of the telescope used to observe the cloud (discussed and quantified in Sec.~\ref{fig: beam resolution study} and Appendix \ref{sec: second moment fwhm dependence} below).

\subsubsection{Gradient (Rotation)-corrected first moment} \label{sec:rotcorr1stmom}
As systematic motions of a cloud can result in a gradient existing in the 1st-moment map \citep{MyersBenson1983, GoldsmithArquilla1985}, this gradient must be subtracted from the 1st-moment map to isolate the turbulent motions of the cloud \citep{FederrathEtAl2016}. We note that on the scale of the cloud, the largest scale (which corresponds to the diameter of the cloud itself) is not considered part of the turbulent motions of the cloud, at least not in the numerical and analytic models of turbulence-regulated star formation \citep{KrumholzMcKee2005,PadoanNordlund2011,HennebelleChabrier2011,FederrathKlessen2012}, for which we are trying to develop a method to measure the velocity dispersion from observations. In order to measure the velocity dispersion (and Mach number) that is used in these analytic models of star formation, we must remove the large-scale gradient from the cloud, because it is not considered to be part of the turbulence of the cloud. However, the large-scale gradient can be, and is often, part of the larger-scale (larger than the cloud) turbulent ISM \citep{MieschBally1994,KleinerDickman1984,KleinerDickman1985,DickmanKleiner1985,RoyJoncas1985,OssenkopfMacLow2002}.

The rotation of the cloud about the $z$-axis causes the 1st-moment map viewed along the $x$ or $y$ axes to exhibit such a gradient. Determining this gradient is advantageous, as it allows for properties of the cloud's motion to be deduced \citep[including its angular momentum; see e.g.,][]{BurkertBodenheimer2000}, and allows for the turbulent motions to be isolated \citep{FederrathEtAl2016}.

We determine the gradient of the cloud's velocity map by porting the \texttt{planefit.pro} IDL function written by H.~T.~Freudenreich in 1993 to Python. This function uses least-squares fitting of a plane of the form
\begin{equation}
\label{eq planefit}
\mathrm{grad} = a + bx + cy
\end{equation}
to a set of points of the form $(x,y,\mathrm{grad}=v_z)$, where $v_z$ is the value of the 1st-moment map at each point $(x,y)$ in the plane of the sky, and $a$, $b$, $c$ are fit parameters. This gradient represents the rotational velocity or large-scale shear contribution to the 1st-moment map of the cloud as viewed along a LOS.

The rotation-corrected 1st-moment map is simply the difference between the 1st-moment map and the gradient fit to the 1st-moment map, given by
\begin{equation}
\label{eq gradient subtracted}
(\langle v_\mathrm{los} \rangle - \mathrm{grad})_p = \langle v_\mathrm{los} \rangle _p - \mathrm{grad} _{\mathrm{los} \> p},
\end{equation}
where $\mathrm{grad} _{\mathrm{los} \> p}$ is the value of the gradient fit to the 1st-moment map as viewed along the LOS axis $\mathrm{los}=[x,y,z]$ at pixel $p$.

We calculate the standard deviation of the rotation-corrected 1st-moment map, $\siggradKD{\mathrm{los}}$, as a quantifier of the cloud turbulence. This statistic measures the velocity dispersion in the visible plane of the cloud and is corrected for the contributions from the systematic motion (rotation) of the cloud. Although this statistic includes variation of the LOS velocities across the visible plane, it does not account for variation of the LOS velocities along the LOS.

\subsubsection{Parent and gradient (rotation)-corrected parent velocity dispersion}
\label{sec parent dispersion def}

Even in a cloud without systematic motion, the spatial mean of the 2nd moment is not a perfect measure of turbulence because information regarding the LOS velocity fluctuations in the plane of the sky is lost due to the spatial average. A similar limitation applies to the standard deviation of the 1st-moment map, which does not contain LOS velocity fluctuations along the LOS. One could therefore examine the parent velocity dispersion, which is a combination of the 1st- and 2nd-moment dispersions, as in \citet{DickmanKleiner1985}, as a potential turbulence quantifier;
\begin{equation}
\label{eq sigma p observable}
\sigpKD{\mathrm{los}}^2 = \sigtwoKD{\mathrm{los}}^2 + \sigKD{\mathrm{los}}^2.
\end{equation}
The 3D parent velocity dispersion can be found by summing the parent velocity dispersions measured along the three Cartesian LOS in quadrature. \citet{DickmanKleiner1985} showed that this statistic is equal to the total 3D velocity dispersion. However, this is not a measurable statistic in observations, where one has access to only one Cartesian LOS, and thus, one would normally assume isotropic turbulence and hence multiply $\sigpKD{\mathrm{los}}$ (from the one LOS that can be measured in an observation) by $\sqrt{3}$ to get the 3D parent velocity dispersion. The main problem with this statistic is that it can include contributions from rotation, depending on the orientation of the cloud with respect to the observer. Thus, without a correction for rotation, $\sigpKD{\mathrm{los}}$ will be different when measured along different LOS, and hence, the assumption of isotropy will fail to recover the intrinsic 3D velocity dispersion.

In order to account for rotational contributions, we modify Eq.~(\ref{eq sigma p observable}) such that it becomes the gradient-corrected parent velocity dispersion, $\sigpgradKD{\mathrm{los}}$. This is defined as
\begin{equation}
\label{eq sigma p corrected observable}
\sigpgradKD{\mathrm{los}}^2 = \sigtwoKD{\mathrm{los}}^2 + \siggradKD{\mathrm{los}}^2.
\end{equation}
This statistic is the most promising turbulence quantifier, because it includes turbulent LOS velocity fluctuations both along the given LOS (from $\sigtwoKD{\mathrm{los}}$) and across the plane of the sky (from $\siggradKD{\mathrm{los}}$), while removing contributions from large-scale systematic motions of the cloud.

\section{Results \label{sec: results}}

\subsection{Moment maps \label{sec: moment maps}}

\begin{figure*}
\centering
\includegraphics[width = \textwidth]{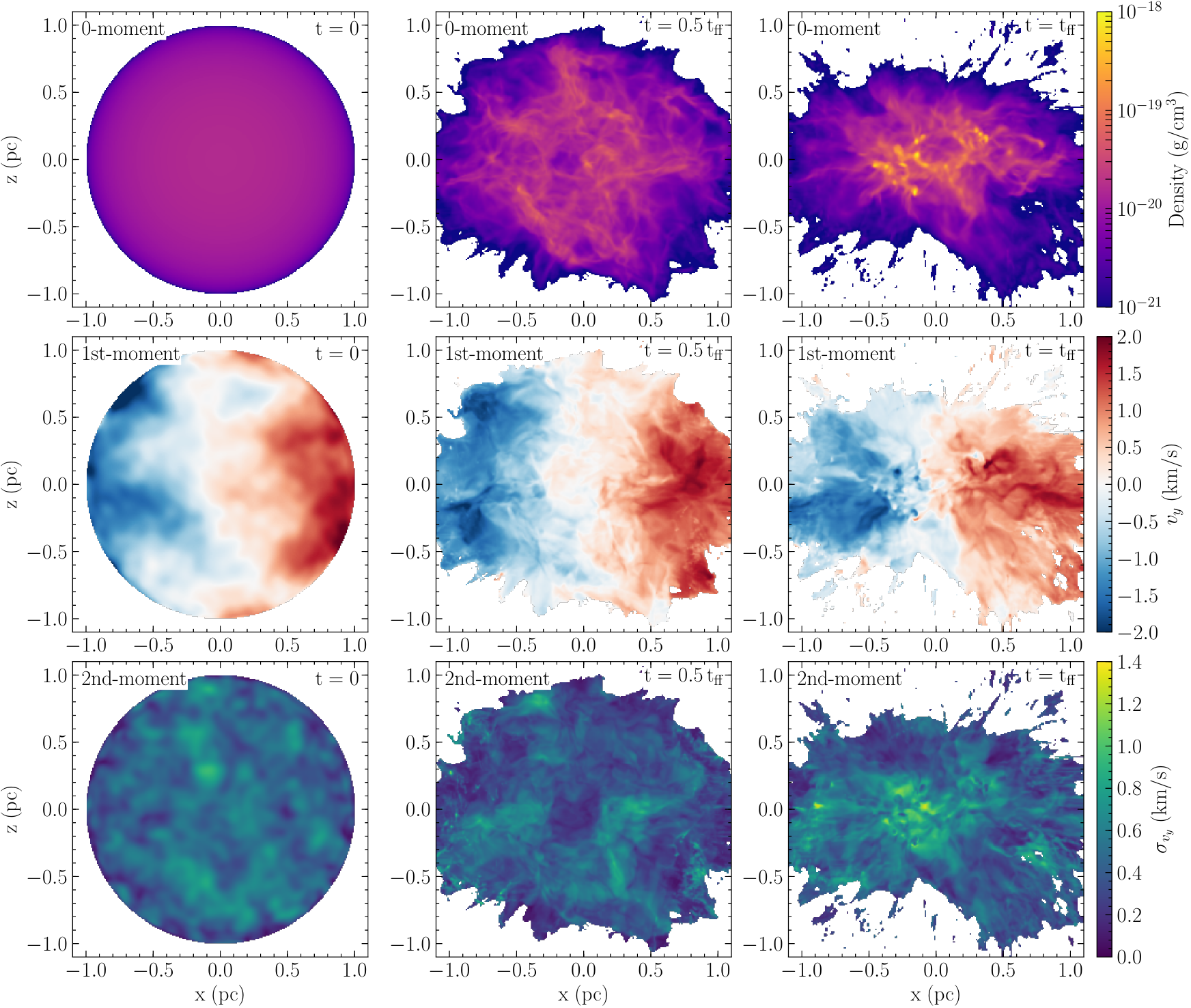}
\caption{Zero-moment maps (1st row), 1st-moment maps (2nd row), and 2nd-moment maps (3rd row) along the $y$-axis LOS for early ($t=0$; left), intermediate ($t=0.5\,\tff$; middle) and late ($t=\tff$; right) simulation times. The column density (1st row) shows how the density structure evolves from initially uniform density to the formation of shocks, filaments and dense cores at late times. The rotation of the cloud is clearly visible in the 1st-moment map (2nd row). The 2nd-moment map (3rd row) is often used to quantify the velocity dispersion. However, we also investigate how to use the 1st-moment map, or a combination of the 1st- and 2nd- moment maps, to measure the velocity dispersion.}
\label{fig: moment maps}
\end{figure*}
    
The structure and physical properties of a cloud can be quantified via its moment maps. Fig.~\ref{fig: moment maps} shows maps of the zero moment (top), first moment (middle), and second moment (bottom) at three different evolutionary stages of the simulated cloud, i.e., at $t=0$ (left), $0.5\,\tff$ (middle), and $\tff$ (right). The zero-moment maps show how the turbulence leads to the formation of shocks and filaments early on, and that at $t=\tff$, a significant fraction of the cloud has collapsed, as indicated by the dense cores towards the centre regions of the cloud. The cloud is also beginning to noticeably contract along the $z$-axis, as expected due to rotation along the $z$ axis \citep[e.g.,][]{MatsumotoHanawaNakamura1997}.
    
The rotation of the cloud is best seen in the velocity (1st-moment) map (2nd row of Fig.~\ref{fig: moment maps}), where a clear velocity gradient is visible. This gradient is of similar magnitude and direction at all times in the simulation. This is consistent with the rotation of the cloud imposed as an initial condition and is similar to gradients in the velocity maps of rotating molecular clouds simulated in e.g., \citet{BurkertBodenheimer2000}. Here we show the LOS along the $y$-axis, but a similar gradient is observed along the $x$ LOS, as a trivial consequence of the rotational symmetry of the cloud around the rotation axis ($z$-axis). In contrast, the velocity maps along the $z$-axis do not exhibit a gradient, which we expect because the cloud is rotating about the $z$-axis (see the 2nd row of Fig.~\ref{fig: moment maps z-axis}).
    
The 2nd-moment maps shown in the 3rd row of Fig.~\ref{fig: moment maps} quantify the LOS variation of velocities (Sec.~\ref{sec:secondmomdef}). The 2nd-moment maps are similar for the two earlier times, but show an increase in the LOS velocity variation towards the centre of the cloud at the later time, which is due to the collapse of the cloud along the rotation axis.

\subsection{Gradient correction \label{sec: gradients and first moments y-axis}}
    
\begin{figure*}
\centering
\includegraphics[width=\textwidth]{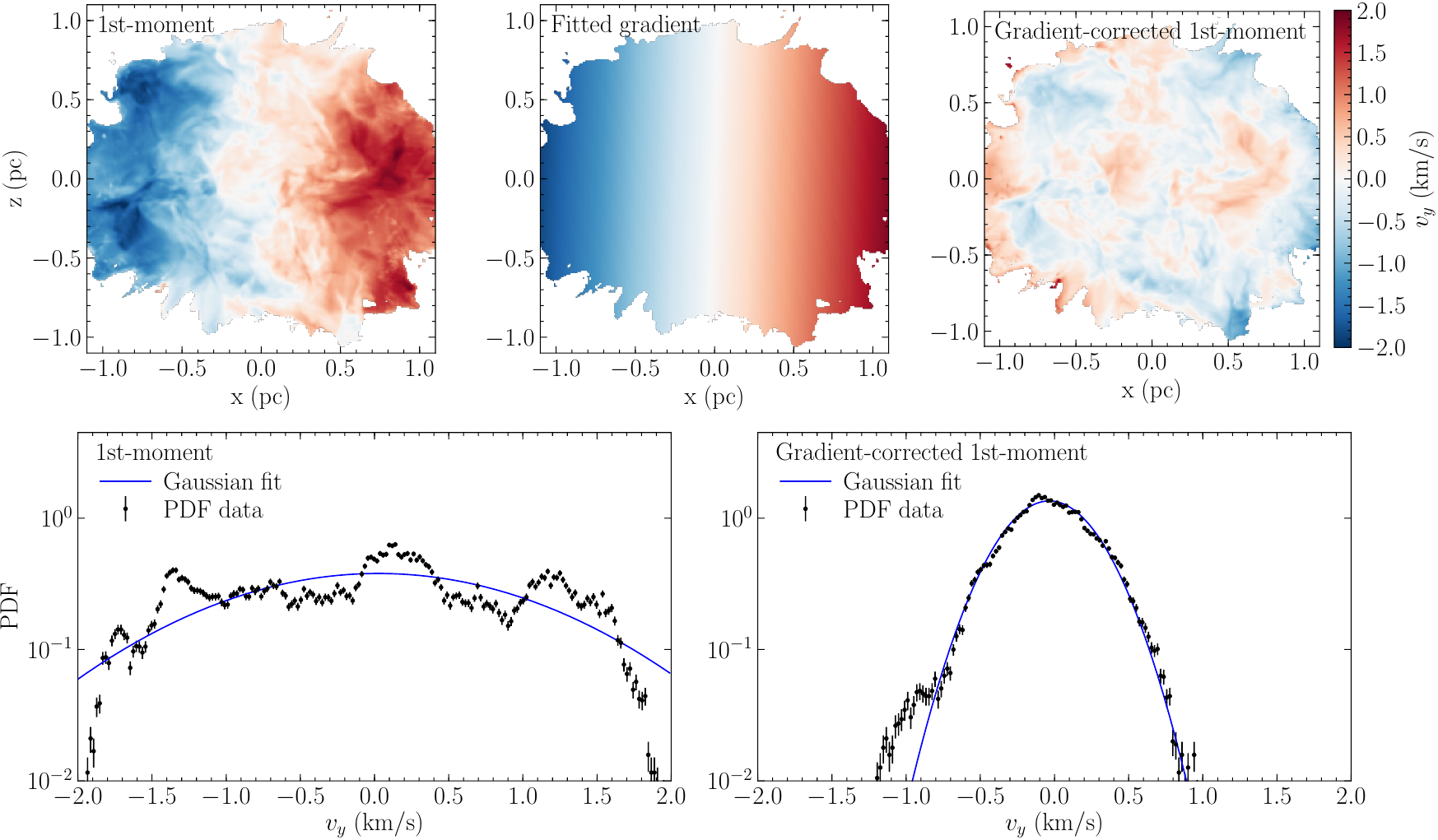}
\caption{Method for subtracting systematic motions (large-scale rotation or shear). Top left panel: original 1st-moment map along the $y$-axis at $t = 0.5 \tff$, showing the rotation of the cloud. Top middle panel: gradient fitted to the 1st-moment map via Eq.~(\ref{eq planefit}). Top right panel: gradient (rotation)-corrected 1st-moment map. The bottom panels show the PDFs of the 1st-moment map before (left) and after (right) the fitted gradient was subtracted, along with Gaussian distributions fitted to each PDF, giving $\sigma_v = 1.05 \pm 0.04\,\km\,\s^{-1}$ and $\sigma_v = 0.295 \pm 0.002\,\km\,\s^{-1}$ before and after gradient subtraction, respectively. Before subtracting the gradient (rotation), the PDF has strong non-Gaussian features (multiple peaks and extended wings), which are removed after the gradient is subtracted, and the resulting PDF is consistent with a near Gaussian PDF, expected for a turbulent medium.}
\label{fig: gradients and first moments y axis}
\end{figure*}

\subsubsection{First-moment maps}
Systematic motions such as large-scale rotation and shear can often be seen in the velocity (1st-moment) maps of molecular clouds \citep{FederrathEtAl2016,ShardaEtAl2018,ShardaEtAl2019,MenonEtAl2021}. However, such large-scale motions should not be considered turbulent motions, at least not on the integral scale (diameter) of the cloud itself. Here we develop a method for measuring the turbulence of a cloud using the 1st-moment map corrected for contributions from such systematic motions. The top left panel of Fig.~\ref{fig: gradients and first moments y axis} shows the 1st-moment map as viewed along the $y$-axis for $t=0.5\,\tff$ in the simulation. We clearly see the large-scale rotation in the map as a gradient. To isolate the turbulent motions, we can subtract a fitted gradient, shown in the top middle panel of Fig.~\ref{fig: gradients and first moments y axis}, from the 1st-moment map \citep{FederrathEtAl2016}.
        
The gradient fit to the 1st-moment map is consistent with the cloud rotating about the $z$-axis. Thus, the fitting method described in Sec.~\ref{sec:rotcorr1stmom} automatically picks up the intrinsic rotation axis. The subtraction of this gradient results in the gradient-corrected 1st-moment map, shown in the top right panel of Fig.~\ref{fig: gradients and first moments y axis}. After subtracting the fitted gradient from the 1st-moment map, it no longer contains a visible gradient, suggesting that we have successfully removed the contribution of the large-scale rotation, and what remains are primarily the turbulent motions of the cloud.
        
The gradient-subtraction method works equally well for any LOS. For example, if we looked along the rotation axis ($z$-axis), there would be no visible gradient in the 1st-moment map, so the fitted gradient would be negligible (near zero), and the gradient-corrected 1st-moment map would be practically identical to the original 1st-moment map (see Fig.~\ref{fig: gradients and first moments z axis}).

\subsubsection{Velocity PDFs}
In order to demonstrate that the gradient-subtraction technique does indeed correct for rotation and isolate the turbulent motions, we show the velocity probability distribution functions (PDFs) of the 1st-moment maps before and after subtracting the gradient in the two bottom panels of Fig.~\ref{fig: gradients and first moments y axis}. The PDF before gradient subtraction does not resemble a Gaussian distribution and has a wide, double-peak contribution from the rotation of the cloud. In contrast, the velocity PDF after gradient subtraction closely resembles a Gaussian distribution, which is the expected distribution for a purely turbulent medium \citep{Klessen2000,Federrath2013}. Note that some deviations from a perfect Gaussian are expected due to intermittent events in the turbulent flow, especially in the wings of the PDF \citep{FederrathEtAl2016}.
        
Before gradient subtraction, the velocity PDF was significantly wider (with $\sigma_v = 1.05\pm0.04\,\km\,\s^{-1}$) than after gradient subtraction ($\sigma_v = 0.295\pm0.002\,\km\,\s^{-1}$). This is due to the rotation of the cloud, increasing the range of velocities of the gas and the dispersion of these velocities. This illustrates the need for contributions from non-turbulent motions to be subtracted in order to isolate the turbulence of a cloud.

\subsection{Time evolution} \label{sec: tevol}

\begin{figure*}
\centering
\includegraphics[width=\textwidth]{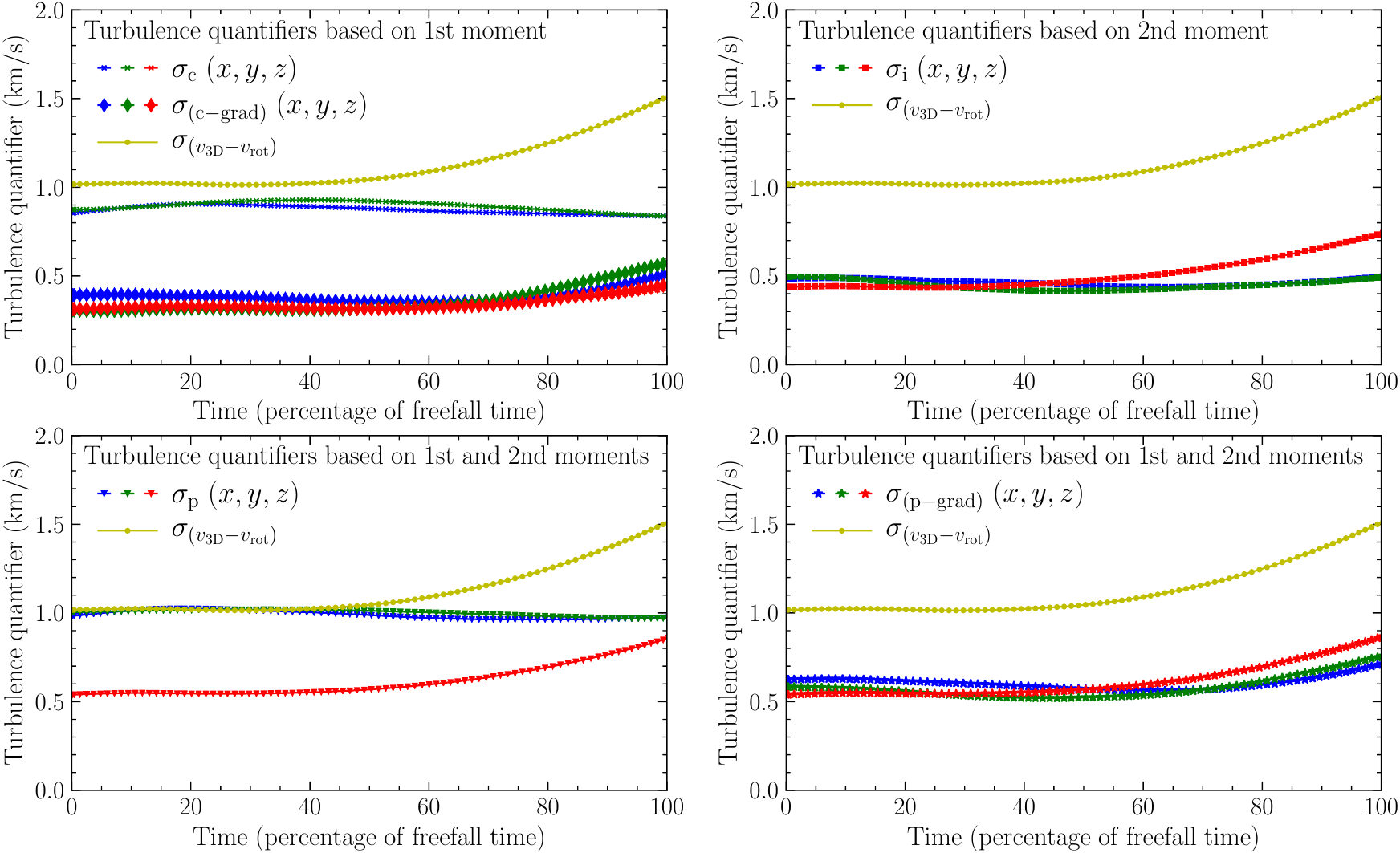}
\caption{Various velocity statistics relevant to the turbulence measured as a function of time. Top left panel: standard deviation of the 1st-moment map before (crosses) and after (diamonds) gradient subtraction as viewed along the $x$ (blue), $y$ (green) and $z$ (red) axes, together with the true 3D turbulent dispersion (yellow circles). Top right, bottom left and bottom right panel: same as top left panel, but for the spatial mean of the 2nd-moment map, the parent velocity dispersion and the gradient-corrected parent velocity dispersion, respectively. The standard deviation of the gradient-subtracted 1st moment, the spatial mean of the 2nd moment and the gradient-corrected parent velocity dispersion do not show a strong dependence on the LOS direction, and appear to follow a similar trend to the 3D turbulent dispersion, all slightly increasing over time as the turbulence feeds from the gravitational collapse of the cloud. On the other hand, any moment-based statistic that does not correct for the gradient in the 1st-moment map is anisotropic and therefore not suitable to recover the 3D dispersion from a PPV observation.}
\label{fig: statistics as a function of time}
\end{figure*}
    
To quantify the correlations between the `true' turbulence and the observable statistics, we calculate the standard deviation of the gradient-corrected 1st-moment map, the spatial mean of the 2nd-moment map, the parent velocity dispersion and the gradient-corrected parent velocity dispersion, and compare them with the rotation-corrected 3D velocity dispersion. We do this for each time-step over the full simulation time, i.e., from 0 to $1$ freefall time, and for LOS along the $x$, $y$, and $z$ axes, separately.

In the absence of driving, we expect the turbulence to decay over time at a rate related to the turbulent crossing time \citep{MacLowEtAl1998,StoneOstrikerGammie1998}. However, in Fig.~\ref{fig: statistics as a function of time}, we see that the 3D turbulent velocity dispersion (yellow circles) remains fairly constant at a level of about $1\,\km\,\s^{-1}$ until roughly 50\% of the freefall time, after which it increases to approximately $1.8\,\km\,\s^{-1}$ at $t=\tff$. This is because turbulent motions are maintained and driven by the gravitational collapse of the cloud \citep{SurEtAl2010,FederrathSurSchleicherBanerjeeKlessen2011,ShardaEtAl2021}.

\begin{figure*}
\centering
\includegraphics[width=\textwidth]{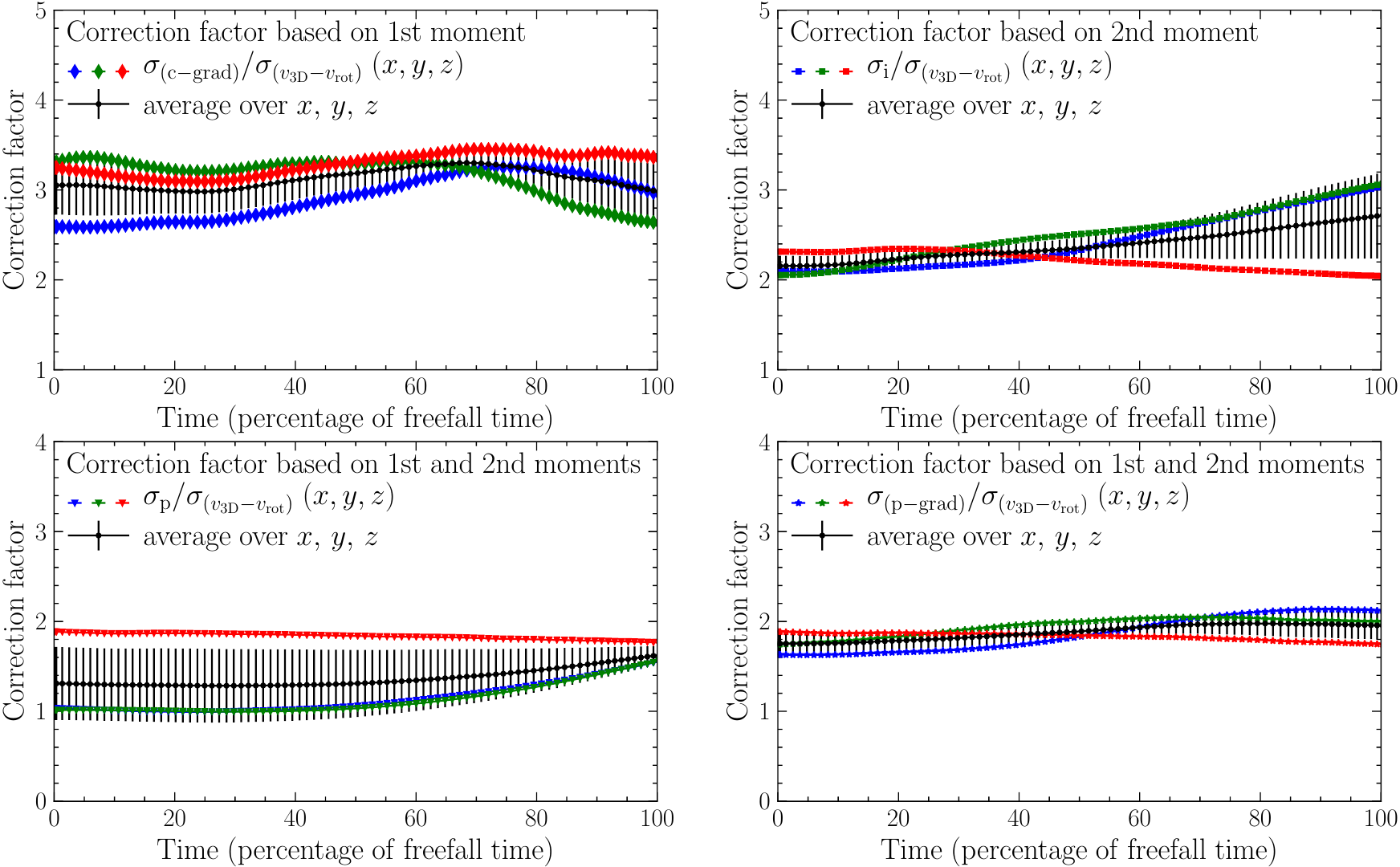}
\caption{Factors by which the standard deviation of the gradient-corrected 1st-moment map (top left panel), the spatial mean of the 2nd-moment map (top right panel), the parent velocity dispersion (bottom left panel) and the gradient-corrected parent velocity dispersion (bottom right panel) can be multiplied to recover the 3D turbulent velocity dispersion of a cloud. The correction factors appear to be relatively independent of the LOS axis and are relatively constant over time for each statistic, except for the parent velocity dispersion, which shows a clear dependence on the LOS axis. The average correction factor over the three LOS is shown as the solid black line with error bars indicating the 1-sigma variations over the three LOS. The gradient-correct parent velocity dispersion (bottom right panel) is clearly the least time-dependent and the least LOS-dependent statistic of all the moment-based methods tested here.}
\label{fig: correction factors}
\end{figure*}

The time evolution of the standard deviation of the 1st moment is shown in the top left-hand panel of Fig.~\ref{fig: statistics as a function of time}. We distinguish the 1st moments before (crosses) and after (diamonds) gradient subtraction (as per Fig.~\ref{fig: gradients and first moments y axis}). The standard deviations along the $x$ (blue) and $y$ (green) axes after gradient subtraction are consistently lower than before the subtraction, as anticipated due to the subtraction removing contributions from the rotation of the cloud. The standard deviations along the $z$-axis (red) are practically identical before and after gradient subtraction (the red crosses are behind the red diamonds in Fig.~\ref{fig: gradients and first moments y axis}, top left panel), because the contribution of rotation to the dispersion is negligible when viewed along the rotation axis (cf., Fig.~\ref{fig: gradients and first moments z axis}). Thus, after gradient subtraction, the standard deviations are similar along the $x$, $y$ and $z$ axes. This is a highly desirable feature of this method, as it shows that irrespective of how the cloud is oriented with respect to the LOS and its rotation axis, we can always correct for systematic motions by removing the fitted large-scale gradient from the 1st-moment map, in order to isolate the turbulent motions in the cloud. Moreover, we see that the time evolution of the gradient-corrected dispersion is similar in shape to the intrinsic 3D turbulent dispersion, offset by a factor (to be determined and discussed in Sec.~\ref{sec: inferring turbulence}), for all times up to $1\,\tff$.
    
The top right-hand panel of Fig.~\ref{fig: statistics as a function of time} shows the same as the top left-hand panel, but for the spatial mean of the 2nd moment instead of the standard deviation of the 1st moment. We see that the spatial mean of the 2nd-moment map is relatively independent of the LOS axis. At first, this result may seem surprising, because the 2nd moment does not correct for the contribution of rotation of the cloud, so we would have expected this statistic to be smaller along the $z$-axis (i.e., no effect of rotation) than along the $x$ and $y$ axes. However, the reason for the 2nd moment not showing a strong dependence on the rotation is that the spatial resolution used in the synthetic observations shown here is relatively high (i.e., it is the intrinsic simulation resolution). For a synthetic observation of the same cloud, but with a large telescope beam (i.e., low spatial resolution), we do find that the 2nd moment increases as expected, which is quantified and discussed in Sec.~\ref{sec: effects of resolution} and Appendix Fig.~\ref{fig: second moment vs fwhm}. After about 50\% of the freefall time, the $z$-direction 2nd moment starts increasing, overtaking the $x$- and $y$-direction 2nd moment. This is because of the collapse of the cloud and the gas contracting more in the $z$ direction, i.e., along the rotation axis, as discussed above in Sec.~\ref{sec: moment maps}.
    
The time evolution of the parent velocity dispersion before and after gradient correction is displayed in the bottom left and right panels of Fig.~\ref{fig: statistics as a function of time}, respectively. In parallel with the 1st moment, the parent velocity dispersion is anisotropic (different in $z$ as compared to $x$ and $y$ LOS) before gradient correction and does not follow the same trend as the intrinsic turbulence. The anisotropy of the standard deviation of the first moment is the major contributor to the LOS dependence of the parent velocity dispersion. This motivates us to use the gradient-corrected first moment instead of the uncorrected first moment to construct the gradient-corrected parent velocity dispersion (shown in the bottom right-hand panel of Fig.~\ref{fig: statistics as a function of time}). This new statistic is nearly independent of the LOS axis and its dependence on time has a similar shape to the true turbulence, making it a promising candidate for a turbulence quantifier.

\section{Inferring the 3D turbulent dispersion from PPV moment-based statistics} \label{sec: inferring turbulence}

In order to recover the 3D turbulent dispersion from any of the moment-based statistics shown in Fig.~\ref{fig: statistics as a function of time} we can multiply the values of these statistics along a LOS by a correction factor. These correction factors have been calculated as a function of time and are displayed in Fig.~\ref{fig: correction factors}.

We see in Fig.~\ref{fig: correction factors} that the correction factors for all statistics are relatively constant in time until about 50\% freefall time. They also appear to be independent of the LOS in all cases except for the uncorrected parent velocity dispersion (bottom left panel of Fig.~\ref{fig: correction factors}). This implies that we can find a LOS-independent correction factor for each statistic that can be used to recover the 3D turbulent velocity dispersion from any LOS, except for the uncorrected parent velocity dispersion, which is dependent on LOS. We calculated this correction factor as an average over the 3~LOS and time. We find correction factors of $3.1 \pm 0.6$ for the standard deviation of the gradient-corrected 1st-moment map, $2.4 \pm 0.3$ for the spatial mean of the 2nd-moment map, and $1.9 \pm 0.2$ for the gradient-corrected parent velocity dispersion.
        
These correction factors for the gradient-corrected 1st moment and the spatial mean of the 2nd moment are greater than the $\sqrt{3}$ factor expected for isotropic turbulence \citep{McKeeOstriker2007}, which means that those statistics do not contain all turbulent velocity fluctuations of the LOS velocity. The main reason for this is that the gradient-correct 1st moment does not include the dispersion along the LOS, only in the plane of the sky, and the mean of the 2nd moment does not fully account for variations in the plane of the sky due to the spatial average. On the other hand, the correction factor for the gradient-corrected parent velocity dispersion is close to the predicted $\sqrt{3}$ factor, showing that $\sigpgradKD{\mathrm{los}}$ contains all turbulent velocity fluctuations for the given LOS velocity component, and corrects for large-scale gradients, such that it is largely independent of the LOS orientation towards the cloud (see bottom right-hand panel of Fig.~\ref{fig: correction factors}).

\section{Effects of the telescope beam resolution} \label{sec: effects of resolution}
\begin{figure}
    \centering
    \includegraphics[width=\linewidth]{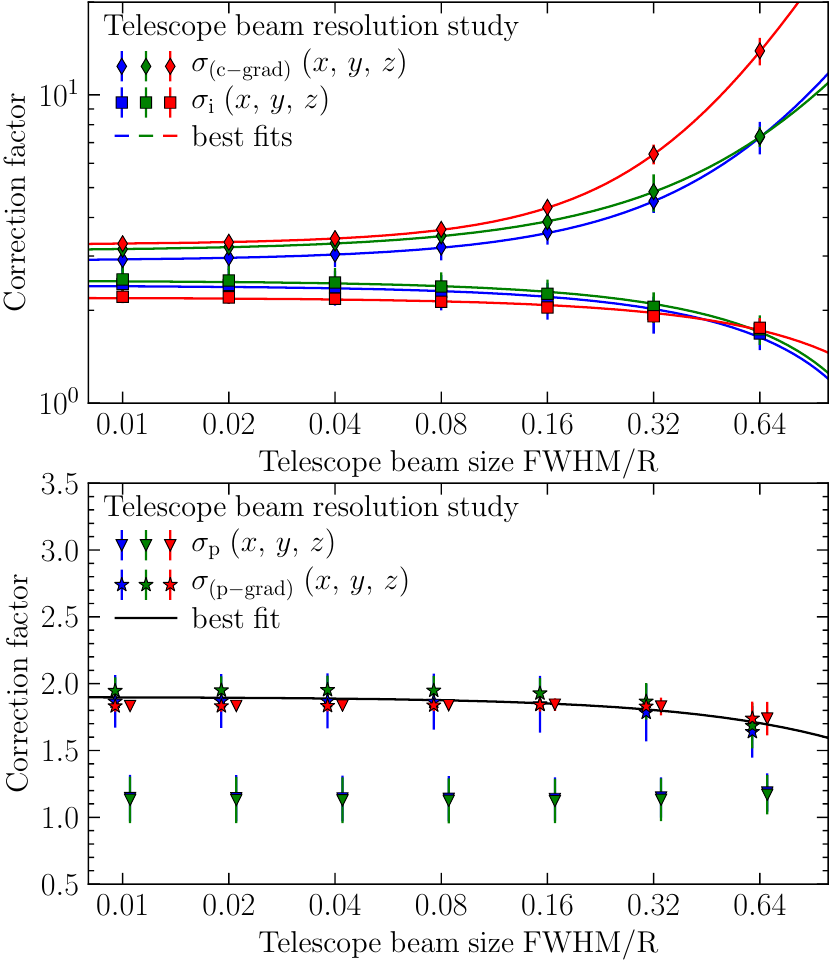}
    \caption{Top panel: time-averaged correction factors needed to recover the 3D turbulent velocity dispersion of a cloud from the standard deviation of the gradient-corrected 1st-moment map (diamonds) or from the spatial mean of the 2nd-moment map (squares), as viewed along the $x$ (blue), $y$ (green) and $z$ (red) axes, as a function of telescope beam FWHM in units of the cloud radius $R$. A quadratic function is fitted to the correction factors of the standard deviation of the gradient-corrected 1st moment and a linear function is fitted to the correction factors of the spatial mean of the 2nd moment. Bottom panel: same as top panel, but for the parent velocity dispersion before (triangles) and after (stars) gradient correction. The time-averaged correction factors for the gradient-corrected parent velocity dispersion measured along the three LOS axes agree within their 1-sigma uncertainties at each given telescope beam FWHM, so a single linear function is fitted to the corresponding correction factors for all LOS. The fit parameters for each of the best fits are given in Tab.~\ref{tab: lines of best fit}.}
    \label{fig: beam resolution study}
\end{figure}

In order to make the correction factors derived in the previous section practically applicable and versatile with respect to the telescope beam resolution, we investigate the effect of the telescope beam size on the observed statistics and correction factors necessary to recover the turbulence. To do this, we decrease the resolution of the synthetic observations of the simulated cloud by applying a Gaussian smoothing function to each moment map. We then repeat the analysis for these beam-smeared maps as before. The correction factors for these statistics were then averaged over time to obtain the correction factors as a function of spatial telescope resolution.

The correction factors based on the gradient-corrected 1st moment, the 2nd moment, and the uncorrected and gradient-corrected parent velocity dispersion are shown as a function of telescope beam resolution in Fig.~\ref{fig: beam resolution study}. We see that the correction factors increase with increasing beam size for the gradient-corrected 1st moment, but decrease with increasing beam size for the 2nd moment (see top panel). We expect an increasing correction factor for the gradient-corrected 1st moment as some of the velocity fluctuations in the plane of the sky are smoothed by the Gaussian smoothing function and hence larger beam widths require higher correction factors to recover the 3D dispersion. Conversely, the 2nd moment increases with increasing beam size due to `beam smearing' (see Sec.~\ref{sec: second moment fwhm dependence}) \citep[also known in the galaxy community; e.g.,][]{VaridelEtAl2016,FederrathEtAl2017,ZhouEtAl2017}, and thus, the respective correction factor decreases with increasing telescope beam size.

Since the parent velocity dispersions (bottom panel of Fig.~\ref{fig: beam resolution study}) combine the 1st and 2nd moments, their correction factors are less dependent on the telescope beam size than the 1st and 2nd moments. The correction factors for the uncorrected parent velocity dispersion are highly dependent on the LOS axis, with lower correction factors observed for the $x$ and $y$ axes compared to the rotation axis, as discussed earlier. This emphasizes that the uncorrected parent velocity dispersion is an unsuitable measure of turbulence in a cloud with systematic motion, such as rotation. Conversely, the correction factors for the gradient-corrected parent velocity dispersion are nearly independent of the LOS axis, even more so than the gradient-corrected 1st moment or 2nd moment themselves.

The relationship between the correction factors and the beam size is well fit by a quadratic function for the gradient-corrected 1st moment and by a linear function for the 2nd moment and the gradient-corrected parent velocity dispersion. The fit parameters for each statistic are listed in Tab.~\ref{tab: lines of best fit}. We do not provide a fit for the uncorrected parent dispersion due to the strong LOS dependence of this statistic. We find that the best statistic, i.e., the one with the least dependence on LOS direction and the least dependence on telescope resolution is the gradient-corrected parent dispersion, for which we find the correction factor $C_{\mathrm{(p}-\mathrm{grad)}}^\mathrm{any} = (-0.31 \pm 0.25)\,f/R + 1.90 \pm 0.14$, where $f$ is the FWHM beam size and $R$ is the cloud radius, shown as the solid black line in the bottom panel of Fig.~\ref{fig: beam resolution study}. We note that the correction factor is nearly constant at the expected isotropic value of $\sqrt{3}$, for any beam size $f/R \lesssim 0.5$. This means that the beam size correction only starts to play a role when the beam size approaches the size of the cloud itself, while for $f/R \lesssim 0.5$, a constant correction factor of $1.90\pm0.14$ may be used.

\section{Effects of the strength of rotation} \label{sec: effects of strength of rotation}
\begin{figure}
    \centering
    \includegraphics[width=\linewidth]{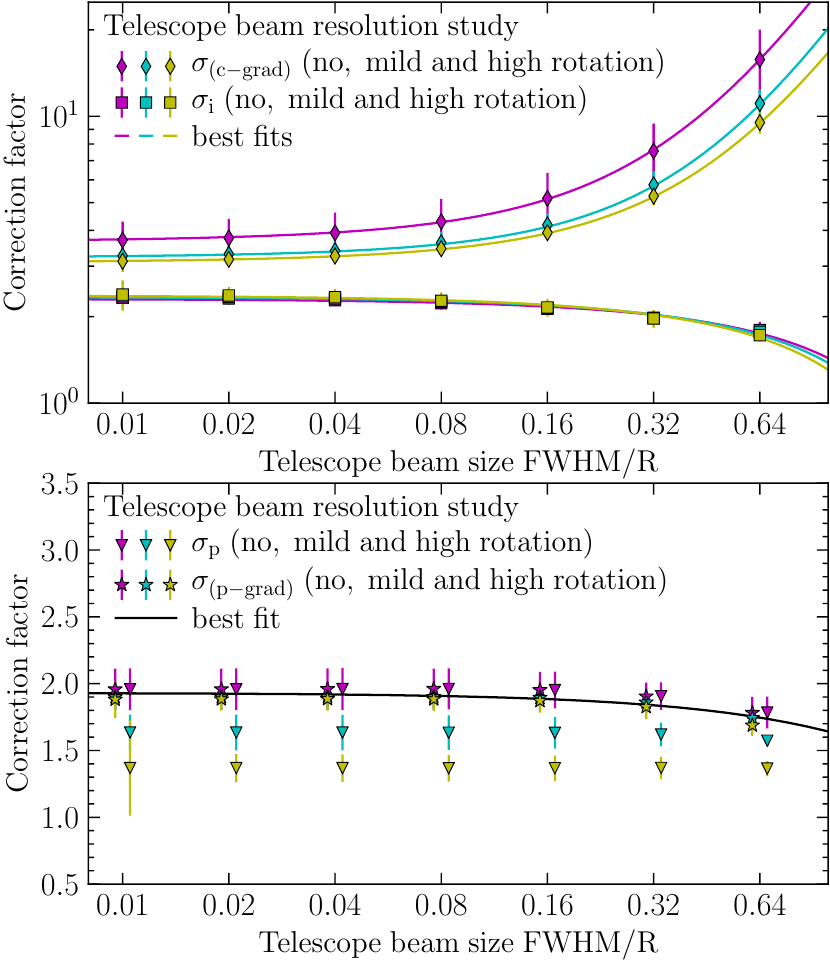}
    \caption{Top panel: LOS- and time-averaged correction factors necessary to recover the turbulent velocity dispersion of a cloud from the standard deviation of the gradient-corrected 1st-moment map (diamonds) and the spatial mean of the 2nd-moment map (squares), as a function of the ratio of telescope beam FWHM to cloud radius $R$, for clouds with no (purple), mild (cyan) and high (yellow) rotation. Quadratic functions are fitted to the correction factors for the standard deviation of the gradient-corrected 1st moment and linear functions are fitted to the correction factors for the spatial mean of the 2nd moment. Bottom panel: same as top panel, but for the parent velocity dispersion before (triangles) and after (stars) gradient correction. A linear function is fitted to the correction factors for the gradient-corrected parent velocity dispersion for all three simulations simultaneously. The fit parameters for each of the best fits are given in Tab.~\ref{tab: lines of best fit}.}
    \label{fig: effects of rotation}
\end{figure}

So far, we have looked at a fixed ratio of rotation and turbulence in the cloud. Here we are adding two simulations, one without rotation and one with mild rotation (ratio of rotation-to-turbulence energy of $E_\mathrm{rot} / E_\mathrm{turb} = 0.35$, compared to the high-rotation simulation with $E_\mathrm{rot} / E_\mathrm{turb} = 1.0$ discussed so far), in order to test the dependence of our results on different levels of rotation. All other parameters are kept the same (see Sec.~\ref{sec: simulations}). To this end, we ran the entire analysis pipeline, including the effects of the telescope beam resolution.

The correction factors for each statistic, averaged over LOS and time, are given for each simulation as a function of telescope beam size in Figure~\ref{fig: effects of rotation}. Consistent with the previous analyses, the gradient-corrected 1st moment and the 2nd moment (top panel) show relatively little dependence on the strength of rotation. The correction factors for each of these statistics agree to within 1-sigma for different levels of rotation. The fitted function parameters are again listed in Tab.~\ref{tab: lines of best fit}.

The correction factors for the uncorrected parent velocity dispersion (bottom panel) are highly dependent on rotation, as expected based on earlier discussion of this quantifier. By contrast, the correction factor for the gradient-corrected parent velocity dispersion is independent of cloud rotation, as anticipated. This makes it an excellent statistic to recover the purely turbulent velocity dispersion of a cloud. A single linear function fits the data from all three simulations very well, and is given by $C_{\mathrm{(p}-\mathrm{grad)}}^\mathrm{any} = (-0.29 \pm 0.26) f/R + (1.93 \pm 0.15)$. This fit is nearly identical to the respective fit from Fig.~\ref{fig: beam resolution study} (bottom panel), and is consistent with a beam-independent correction factor of $1.9\pm0.2$ for $f/R \lesssim 0.5$.

\section{Conclusions} \label{sec: conclusion}

We ran and analysed a numerical simulation of an idealised rotating (around the $z$ axis), collapsing molecular cloud with a prescribed 3D turbulent velocity dispersion. We compute synthetic moment maps of the simulation along the $x$, $y$, and $z$ axes of the cloud, and quantify the spatial mean of the 2nd moment, the standard deviations of the original and gradient-corrected 1st moment, and the original and gradient-corrected parent velocity dispersion. We find that we can recover the intrinsic 3D turbulent velocity dispersion by multiplying different correction factors for each of these statistics. We find that the gradient-corrected parent velocity dispersion, which combines information from the 1st and 2nd moment, provides the most robust reconstruction of the 3D turbulent velocity dispersion, independent of the cloud's rotation, the LOS orientation with respect to the rotation axis, and largely independent of the evolutionary stage of the cloud. We also find that the gradient-corrected parent dispersion is the statistic with the weakest dependence on telescope beam resolution. In the following we list the main results of this study.

\begin{itemize}

\item The zero-moment maps of the cloud at early, intermediate and late times in the simulation show the formation of shocks, filaments and dense pre-stellar cores in the cloud (Fig.~\ref{fig: moment maps}).

\item There is an observable gradient in the 1st-moment map of the cloud along the $x$ and $y$ axes, but not the $z$-axis, which is due to rotation of the cloud about the $z$-axis. We find that the velocity PDF of the 1st moment exhibits strong non-Gaussian features when viewed along the $x$ and $y$ axes. These deviations from a Gaussian distribution are caused by the rotation of the cloud and manifest as the gradient observed in the 1st-moment map.

\item We fit a 2D gradient to the 1st-moment map and subtract it in order to correct for the rotation-induced gradient. This isolates the turbulent motions of the cloud as seen in the gradient-corrected 1st-moment map and its velocity PDF, which is close to a Gaussian distribution, as expected for a purely turbulent medium (Fig.~\ref{fig: gradients and first moments y axis}).

\item The standard deviation of the gradient-corrected 1st-moment map, the spatial mean of the 2nd-moment map, and the gradient-corrected parent velocity dispersion are all nearly independent of the LOS and follow similar trends to the 3D turbulent velocity dispersion of the cloud, which we aim to recover (Fig.~\ref{fig: statistics as a function of time}). This means that we can construct correction factors, i.e., the ratio of the 3D dispersion and either the gradient-corrected 1st moment, the spatial mean of the 2nd moment, or the gradient-corrected parent velocity dispersion, respectively. The uncorrected parent velocity dispersion does not exhibit the necessary properties to construct a correction factor independent of LOS and evolution time.

\item We find the LOS- and time-averaged correction factors to recover the 3D turbulent velocity dispersion are $3.1 \pm 0.6$ for the gradient-corrected 1st moment, $2.4 \pm 0.3$ for the spatial mean of the 2nd-moment map, and $1.9 \pm 0.2$ for the gradient-corrected parent velocity dispersion (Fig.~\ref{fig: correction factors}). The correction factors for the gradient-corrected 1st moment and the 2nd moment are higher than the $\sqrt{3}$ correction factor normally applied based on isotropic turbulence. We attribute this to both these statistics omitting some components of the fluctuations of the LOS velocity: the gradient-corrected 1st moment only contains variations of the LOS velocity in the plane of the sky, but not along the LOS, while the spatial mean of the 2nd moment does not fully account for variations of the LOS velocity in the plane of the sky. The correction factor for the gradient-corrected parent velocity dispersion is much closer to the ideal $\sqrt{3}$ factor, indicating that it is a more accurate measure of the turbulent LOS velocity dispersion.

\item The correction factors for the standard deviation of the gradient-corrected 1st moment and the spatial mean of the 2nd moment are both relatively independent of LOS at high telescope resolutions, but should be used with caution at lower resolutions when quantifying turbulence (Fig.~\ref{fig: beam resolution study}). The correction factor for the gradient-corrected parent velocity dispersion is statistically independent of the LOS, and is only very weakly dependent on telescope beam size.

\item Correction factors for the standard deviation of the gradient-corrected 1st-moment and the spatial mean of the 2nd moment are only weakly dependent on the level of rotation of the cloud. The uncorrected parent velocity dispersion is highly dependent on the level of rotation. By contrast, the correction factor for the gradient-corrected parent velocity dispersion is statistically independent of the level of rotation of the cloud, making it an excellent quantifier of the 3D turbulent velocity dispersion (Fig.~\ref{fig: effects of rotation}).

\item Given a telescope beam FWHM of $f$ relative to the observed cloud radius $R$, we find that the 3D turbulent dispersion can best be recovered via $\sigvtrue = \left[(-0.29\pm0.26)\,f/R + 1.93\pm0.15 \right] \sigpgradKD{\mathrm{los}}$, where $\sigpgradKD{\mathrm{los}}$ is the gradient-correct parent velocity dispersion, defined in Sec.~\ref{sec parent dispersion def}.

\end{itemize}
While the methods and calibrations developed here provide excellent means to reconstruct the 3D turbulent velocity dispersion for a wide range of cloud rotation, independent of the viewing angle, we have not tested other cloud parameters, such as the sonic Mach number, the driving mode of the turbulence, or the magnetic field strength and geometry. Future work should test whether the present calibration holds for different physical and geometrical cloud parameters.

\section*{Acknowledgements}
We thank the two anonymous referees for their comments, which improved this work.
C.~F.~acknowledges funding provided by the Australian Research Council (Future Fellowship FT180100495), and the Australia-Germany Joint Research Cooperation Scheme (UA-DAAD). We further acknowledge high-performance computing resources provided by the Australian National Computational Infrastructure (grant~ek9) in the framework of the National Computational Merit Allocation Scheme and the ANU Merit Allocation Scheme, and by the Leibniz Rechenzentrum and the Gauss Centre for Supercomputing (grant~pr32lo). The simulation software FLASH was in part developed by the DOE-supported Flash Center for Computational Science at the University of Chicago.

\section*{Data availability}
The simulation data in this article will be shared upon reasonable request to Christoph Federrath (\href{mailto:christoph.federrath@anu.edu.au}{christoph.federrath@anu.edu.au}).




\def\rmp{{\sl Rev.~Mod.~Phys.}}

\begin{thebibliography}{}
\makeatletter
\relax
\def\mn@urlcharsother{\let\do\@makeother \do\$\do\&\do\#\do\^\do\_\do\%\do\~}
\def\mn@doi{\begingroup\mn@urlcharsother \@ifnextchar [ {\mn@doi@}
  {\mn@doi@[]}}
\def\mn@doi@[#1]#2{\def\@tempa{#1}\ifx\@tempa\@empty \href
  {http://dx.doi.org/#2} {doi:#2}\else \href {http://dx.doi.org/#2} {#1}\fi
  \endgroup}
\def\mn@eprint#1#2{\mn@eprint@#1:#2::\@nil}
\def\mn@eprint@arXiv#1{\href {http://arxiv.org/abs/#1} {{\tt arXiv:#1}}}
\def\mn@eprint@dblp#1{\href {http://dblp.uni-trier.de/rec/bibtex/#1.xml}
  {dblp:#1}}
\def\mn@eprint@#1:#2:#3:#4\@nil{\def\@tempa {#1}\def\@tempb {#2}\def\@tempc
  {#3}\ifx \@tempc \@empty \let \@tempc \@tempb \let \@tempb \@tempa \fi \ifx
  \@tempb \@empty \def\@tempb {arXiv}\fi \@ifundefined
  {mn@eprint@\@tempb}{\@tempb:\@tempc}{\expandafter \expandafter \csname
  mn@eprint@\@tempb\endcsname \expandafter{\@tempc}}}

\bibitem[\protect\citeauthoryear{{Andr{\'e}}, {Di Francesco}, {Ward-Thompson},
  {Inutsuka}, {Pudritz}  \& {Pineda}}{{Andr{\'e}} et~al.}{2014}]{AndreEtAl2014}
{Andr{\'e}} P.,  {Di Francesco} J.,  {Ward-Thompson} D.,  {Inutsuka} S.-I.,
  {Pudritz} R.~E.,   {Pineda} J.~E.,  2014, in {Beuther} H.,  {Klessen} R.~S.,
  {Dullemond} C.~P.,   {Henning} T.,  eds, Protostars and Planets VI.
  University of Arizona Press, pp 27--51 (\mn@eprint {arXiv} {1312.6232}),
  \mn@doi{10.2458/azu_uapress_9780816531240-ch002}

\bibitem[\protect\citeauthoryear{{Arzoumanian} et~al.,}{{Arzoumanian}
  et~al.}{2011}]{ArzoumanianEtAl2011}
{Arzoumanian} D.,  et~al., 2011, \mn@doi [\aap] {10.1051/0004-6361/201116596},
  \href {http://adsabs.harvard.edu/abs/2011A%26A...529L...6A} {529, L6}

\bibitem[\protect\citeauthoryear{{Berger} \& {Colella}}{{Berger} \&
  {Colella}}{1989}]{BergerColella1989}
{Berger} M.~J.,  {Colella} P.,  1989, \mn@doi [\jcp]
  {10.1016/0021-9991(89)90035-1}, 82, 64

\bibitem[\protect\citeauthoryear{{Brunt} \& {Federrath}}{{Brunt} \&
  {Federrath}}{2014}]{BruntFederrath2014}
{Brunt} C.~M.,  {Federrath} C.,  2014, \mn@doi [\mnras] {10.1093/mnras/stu888},
  \href {http://adsabs.harvard.edu/abs/2014MNRAS.442.1451B} {442, 1451}

\bibitem[\protect\citeauthoryear{{Brunt}, {Federrath}  \& {Price}}{{Brunt}
  et~al.}{2010a}]{BruntFederrathPrice2010a}
{Brunt} C.~M.,  {Federrath} C.,   {Price} D.~J.,  2010a, \mn@doi [\mnras]
  {10.1111/j.1365-2966.2009.16215.x}, \href
  {http://adsabs.harvard.edu/abs/2010MNRAS.403.1507B} {403, 1507}

\bibitem[\protect\citeauthoryear{{Brunt}, {Federrath}  \& {Price}}{{Brunt}
  et~al.}{2010b}]{BruntFederrathPrice2010b}
{Brunt} C.~M.,  {Federrath} C.,   {Price} D.~J.,  2010b, \mn@doi [\mnras]
  {10.1111/j.1745-3933.2010.00858.x}, \href
  {http://adsabs.harvard.edu/abs/2010MNRAS.405L..56B} {405, L56}

\bibitem[\protect\citeauthoryear{{Burkert} \& {Bodenheimer}}{{Burkert} \&
  {Bodenheimer}}{2000}]{BurkertBodenheimer2000}
{Burkert} A.,  {Bodenheimer} P.,  2000, \mn@doi [\apj] {10.1086/317122}, \href
  {http://adsabs.harvard.edu/abs/2000ApJ...543..822B} {543, 822}

\bibitem[\protect\citeauthoryear{{Burkhart} \& {Lazarian}}{{Burkhart} \&
  {Lazarian}}{2012}]{BurkhartLazarian2012}
{Burkhart} B.,  {Lazarian} A.,  2012, \mn@doi [\apjl]
  {10.1088/2041-8205/755/1/L19}, \href
  {http://esoads.eso.org/abs/2012ApJ...755L..19B} {755, L19}

\bibitem[\protect\citeauthoryear{{Dickman} \& {Kleiner}}{{Dickman} \&
  {Kleiner}}{1985}]{DickmanKleiner1985}
{Dickman} R.~L.,  {Kleiner} S.~C.,  1985, \mn@doi [\apj] {10.1086/163391},
  \href {https://ui.adsabs.harvard.edu/abs/1985ApJ...295..479D} {295, 479}

\bibitem[\protect\citeauthoryear{{Dubey} et~al.,}{{Dubey}
  et~al.}{2008}]{DubeyEtAl2008}
{Dubey} A.,  et~al., 2008, in {Pogorelov} N.~V.,  {Audit} E.,   {Zank} G.~P.,
  eds,  Astronomical Society of the Pacific Conference Series Vol. 385,
  Numerical Modeling of Space Plasma Flows. p.~145

\bibitem[\protect\citeauthoryear{{Federrath}}{{Federrath}}{2013}]{Federrath2013}
{Federrath} C.,  2013, \mn@doi [\mnras] {10.1093/mnras/stt1644}, \href
  {http://adsabs.harvard.edu/abs/2013MNRAS.436.1245F} {436, 1245}

\bibitem[\protect\citeauthoryear{{Federrath}}{{Federrath}}{2018}]{Federrath2018}
{Federrath} C.,  2018, \mn@doi [Physics Today] {10.1063/PT.3.3947}, \href
  {http://adsabs.harvard.edu/abs/2018PhT....71f..38F} {71, 38}

\bibitem[\protect\citeauthoryear{{Federrath} \& {Klessen}}{{Federrath} \&
  {Klessen}}{2012}]{FederrathKlessen2012}
{Federrath} C.,  {Klessen} R.~S.,  2012, \mn@doi [\apj]
  {10.1088/0004-637X/761/2/156}, \href
  {http://adsabs.harvard.edu/abs/2012ApJ...761..156F} {761, 156}

\bibitem[\protect\citeauthoryear{{Federrath}, {Roman-Duval}, {Klessen},
  {Schmidt}  \& {Mac Low}}{{Federrath}
  et~al.}{2010a}]{FederrathDuvalKlessenSchmidtMacLow2010}
{Federrath} C.,  {Roman-Duval} J.,  {Klessen} R.~S.,  {Schmidt} W.,   {Mac Low}
  M.,  2010a, \mn@doi [\aap] {10.1051/0004-6361/200912437}, \href
  {http://adsabs.harvard.edu/abs/2010A%26A...512A..81F} {512, A81}

\bibitem[\protect\citeauthoryear{{Federrath}, {Banerjee}, {Clark}  \&
  {Klessen}}{{Federrath} et~al.}{2010b}]{FederrathBanerjeeClarkKlessen2010}
{Federrath} C.,  {Banerjee} R.,  {Clark} P.~C.,   {Klessen} R.~S.,  2010b,
  \mn@doi [\apj] {10.1088/0004-637X/713/1/269}, \href
  {http://cdsads.u-strasbg.fr/abs/2010ApJ...713..269F} {713, 269}

\bibitem[\protect\citeauthoryear{{Federrath}, {Sur}, {Schleicher}, {Banerjee}
  \& {Klessen}}{{Federrath}
  et~al.}{2011}]{FederrathSurSchleicherBanerjeeKlessen2011}
{Federrath} C.,  {Sur} S.,  {Schleicher} D.~R.~G.,  {Banerjee} R.,   {Klessen}
  R.~S.,  2011, \mn@doi [\apj] {10.1088/0004-637X/731/1/62}, \href
  {http://adsabs.harvard.edu/abs/2011ApJ...731...62F} {731, 62}

\bibitem[\protect\citeauthoryear{{Federrath}, {Schr{\"o}n}, {Banerjee}  \&
  {Klessen}}{{Federrath} et~al.}{2014}]{FederrathEtAl2014}
{Federrath} C.,  {Schr{\"o}n} M.,  {Banerjee} R.,   {Klessen} R.~S.,  2014,
  \mn@doi [\apj] {10.1088/0004-637X/790/2/128}, \href
  {http://adsabs.harvard.edu/abs/2014ApJ...790..128F} {790, 128}

\bibitem[\protect\citeauthoryear{{Federrath} et~al.,}{{Federrath}
  et~al.}{2016}]{FederrathEtAl2016}
{Federrath} C.,  et~al., 2016, \mn@doi [\apj] {10.3847/0004-637X/832/2/143},
  \href {http://adsabs.harvard.edu/abs/2016ApJ...832..143F} {832, 143}

\bibitem[\protect\citeauthoryear{{Federrath} et~al.,}{{Federrath}
  et~al.}{2017}]{FederrathEtAl2017}
{Federrath} C.,  et~al., 2017, \mn@doi [\mnras] {10.1093/mnras/stx727}, \href
  {http://adsabs.harvard.edu/abs/2017MNRAS.468.3965F} {468, 3965}

\bibitem[\protect\citeauthoryear{{Federrath}, {Klessen}, {Iapichino}  \&
  {Beattie}}{{Federrath} et~al.}{2021}]{FederrathEtAl2021}
{Federrath} C.,  {Klessen} R.~S.,  {Iapichino} L.,   {Beattie} J.~R.,  2021,
  \mn@doi [Nature Astronomy] {10.1038/s41550-020-01282-z}, \href
  {https://ui.adsabs.harvard.edu/abs/2021NatAs...5..365F} {5, 365}

\bibitem[\protect\citeauthoryear{{Fryxell} et~al.,}{{Fryxell}
  et~al.}{2000}]{FryxellEtAl2000}
{Fryxell} B.,  et~al., 2000, \mn@doi [\apjs] {10.1086/317361}, \href
  {http://cdsads.u-strasbg.fr/abs/2000ApJS..131..273F} {131, 273}

\bibitem[\protect\citeauthoryear{{Gerrard}, {Federrath}  \&
  {Kuruwita}}{{Gerrard} et~al.}{2019}]{GerrardFederrathKuruwita2019}
{Gerrard} I.~A.,  {Federrath} C.,   {Kuruwita} R.,  2019, \mn@doi [\mnras]
  {10.1093/mnras/stz784}, \href
  {https://ui.adsabs.harvard.edu/abs/2019MNRAS.485.5532G} {485, 5532}

\bibitem[\protect\citeauthoryear{{Goldsmith} \& {Arquilla}}{{Goldsmith} \&
  {Arquilla}}{1985}]{GoldsmithArquilla1985}
{Goldsmith} P.~F.,  {Arquilla} R.,  1985, in {Black} D.~C.,  {Matthews} M.~S.,
  eds, Protostars and Planets II. pp 137--149

\bibitem[\protect\citeauthoryear{{Hennebelle} \& {Chabrier}}{{Hennebelle} \&
  {Chabrier}}{2008}]{HennebelleChabrier2008}
{Hennebelle} P.,  {Chabrier} G.,  2008, \mn@doi [\apj] {10.1086/589916}, \href
  {http://cdsads.u-strasbg.fr/abs/2008ApJ...684..395H} {684, 395}

\bibitem[\protect\citeauthoryear{{Hennebelle} \& {Chabrier}}{{Hennebelle} \&
  {Chabrier}}{2009}]{HennebelleChabrier2009}
{Hennebelle} P.,  {Chabrier} G.,  2009, \mn@doi [\apj]
  {10.1088/0004-637X/702/2/1428}, \href
  {http://adsabs.harvard.edu/abs/2009ApJ...702.1428H} {702, 1428}

\bibitem[\protect\citeauthoryear{{Hennebelle} \& {Chabrier}}{{Hennebelle} \&
  {Chabrier}}{2011}]{HennebelleChabrier2011}
{Hennebelle} P.,  {Chabrier} G.,  2011, \mn@doi [\apjl]
  {10.1088/2041-8205/743/2/L29}, \href
  {http://adsabs.harvard.edu/abs/2011ApJ...743L..29H} {743, L29}

\bibitem[\protect\citeauthoryear{{Hennebelle} \& {Falgarone}}{{Hennebelle} \&
  {Falgarone}}{2012}]{HennebelleFalgarone2012}
{Hennebelle} P.,  {Falgarone} E.,  2012, \mn@doi [\aapr]
  {10.1007/s00159-012-0055-y}, \href
  {http://adsabs.harvard.edu/abs/2012A%26ARv..20...55H} {20, 55}

\bibitem[\protect\citeauthoryear{{Heyer} \& {Brunt}}{{Heyer} \&
  {Brunt}}{2004}]{HeyerBrunt2004}
{Heyer} M.~H.,  {Brunt} C.~M.,  2004, \mn@doi [\apjl] {10.1086/425978}, \href
  {http://cdsads.u-strasbg.fr/abs/2004ApJ...615L..45H} {615, L45}

\bibitem[\protect\citeauthoryear{{Hopkins}}{{Hopkins}}{2012}]{Hopkins2012a}
{Hopkins} P.~F.,  2012, \mn@doi [\mnras] {10.1111/j.1365-2966.2012.20730.x},
  \href {http://adsabs.harvard.edu/abs/2012MNRAS.423.2016H} {423, 2016}

\bibitem[\protect\citeauthoryear{{Hopkins}}{{Hopkins}}{2013}]{Hopkins2013IMF}
{Hopkins} P.~F.,  2013, \mn@doi [\mnras] {10.1093/mnras/sts704}, \href
  {http://adsabs.harvard.edu/abs/2013MNRAS.430.1653H} {430, 1653}

\bibitem[\protect\citeauthoryear{{Kainulainen}, {Federrath}  \&
  {Henning}}{{Kainulainen} et~al.}{2014}]{KainulainenFederrathHenning2014}
{Kainulainen} J.,  {Federrath} C.,   {Henning} T.,  2014, \mn@doi [Science]
  {10.1126/science.1248724}, \href
  {http://adsabs.harvard.edu/abs/2014Sci...344..183K} {344, 183}

\bibitem[\protect\citeauthoryear{{Kleiner} \& {Dickman}}{{Kleiner} \&
  {Dickman}}{1984}]{KleinerDickman1984}
{Kleiner} S.~C.,  {Dickman} R.~L.,  1984, \mn@doi [\apj] {10.1086/162593},
  \href {https://ui.adsabs.harvard.edu/abs/1984ApJ...286..255K} {286, 255}

\bibitem[\protect\citeauthoryear{{Kleiner} \& {Dickman}}{{Kleiner} \&
  {Dickman}}{1985}]{KleinerDickman1985}
{Kleiner} S.~C.,  {Dickman} R.~L.,  1985, \mn@doi [\apj] {10.1086/163390},
  \href {https://ui.adsabs.harvard.edu/abs/1985ApJ...295..466K} {295, 466}

\bibitem[\protect\citeauthoryear{{Klessen}}{{Klessen}}{2000}]{Klessen2000}
{Klessen} R.~S.,  2000, \mn@doi [\apj] {10.1086/308854}, \href
  {http://cdsads.u-strasbg.fr/abs/2000ApJ...535..869K} {535, 869}

\bibitem[\protect\citeauthoryear{{Kritsuk}, {Norman}, {Padoan}  \&
  {Wagner}}{{Kritsuk} et~al.}{2007}]{KritsukEtAl2007}
{Kritsuk} A.~G.,  {Norman} M.~L.,  {Padoan} P.,   {Wagner} R.,  2007, \mn@doi
  [\apj] {10.1086/519443}, 665, 416

\bibitem[\protect\citeauthoryear{{Krumholz} \& {McKee}}{{Krumholz} \&
  {McKee}}{2005}]{KrumholzMcKee2005}
{Krumholz} M.~R.,  {McKee} C.~F.,  2005, \mn@doi [\apj] {10.1086/431734}, \href
  {http://cdsads.u-strasbg.fr/abs/2005ApJ...630..250K} {630, 250}

\bibitem[\protect\citeauthoryear{{Kuruwita} \& {Federrath}}{{Kuruwita} \&
  {Federrath}}{2019}]{KuruwitaFederrath2019}
{Kuruwita} R.~L.,  {Federrath} C.,  2019, \mn@doi [\mnras]
  {10.1093/mnras/stz1053}, \href
  {https://ui.adsabs.harvard.edu/abs/2019MNRAS.486.3647K} {486, 3647}

\bibitem[\protect\citeauthoryear{{Kuruwita}, {Federrath}  \&
  {Haugb{\o}lle}}{{Kuruwita} et~al.}{2020}]{KuruwitaFederrathHaugboelle2020}
{Kuruwita} R.~L.,  {Federrath} C.,   {Haugb{\o}lle} T.,  2020, \mn@doi [\aap]
  {10.1051/0004-6361/202038181}, \href
  {https://ui.adsabs.harvard.edu/abs/2020A&A...641A..59K} {641, A59}

\bibitem[\protect\citeauthoryear{{Larson}}{{Larson}}{1981}]{Larson1981}
{Larson} R.~B.,  1981, \mnras, \href
  {http://cdsads.u-strasbg.fr/abs/1981MNRAS.194..809L} {194, 809}

\bibitem[\protect\citeauthoryear{{Mac Low} \& {Klessen}}{{Mac Low} \&
  {Klessen}}{2004}]{MacLowKlessen2004}
{Mac Low} M.-M.,  {Klessen} R.~S.,  2004, \rmp, 76, 125

\bibitem[\protect\citeauthoryear{{Mac Low}, {Klessen}, {Burkert}  \&
  {Smith}}{{Mac Low} et~al.}{1998}]{MacLowEtAl1998}
{Mac Low} M.-M.,  {Klessen} R.~S.,  {Burkert} A.,   {Smith} M.~D.,  1998, \prl,
  \href {http://cdsads.u-strasbg.fr/abs/1998PhRvL..80.2754M} {80, 2754}

\bibitem[\protect\citeauthoryear{{Matsumoto}, {Hanawa}  \&
  {Nakamura}}{{Matsumoto} et~al.}{1997}]{MatsumotoHanawaNakamura1997}
{Matsumoto} T.,  {Hanawa} T.,   {Nakamura} F.,  1997, \mn@doi [\apj]
  {10.1086/303822}, \href
  {https://ui.adsabs.harvard.edu/abs/1997ApJ...478..569M} {478, 569}

\bibitem[\protect\citeauthoryear{{McKee} \& {Ostriker}}{{McKee} \&
  {Ostriker}}{2007}]{McKeeOstriker2007}
{McKee} C.~F.,  {Ostriker} E.~C.,  2007, \mn@doi [\araa]
  {10.1146/annurev.astro.45.051806.110602}, \href
  {http://cdsads.u-strasbg.fr/abs/2007ARA%26A..45..565M} {45, 565}

\bibitem[\protect\citeauthoryear{{Menon}, {Federrath}, {Klaassen}, {Kuiper}  \&
  {Reiter}}{{Menon} et~al.}{2021}]{MenonEtAl2021}
{Menon} S.~H.,  {Federrath} C.,  {Klaassen} P.,  {Kuiper} R.,   {Reiter} M.,
  2021, \mn@doi [\mnras] {10.1093/mnras/staa3271}, \href
  {https://ui.adsabs.harvard.edu/abs/2021MNRAS.500.1721M} {500, 1721}

\bibitem[\protect\citeauthoryear{{Miesch} \& {Bally}}{{Miesch} \&
  {Bally}}{1994}]{MieschBally1994}
{Miesch} M.~S.,  {Bally} J.,  1994, \mn@doi [\apj] {10.1086/174352}, \href
  {http://adsabs.harvard.edu/abs/1994ApJ...429..645M} {429, 645}

\bibitem[\protect\citeauthoryear{{Myers} \& {Benson}}{{Myers} \&
  {Benson}}{1983}]{MyersBenson1983}
{Myers} P.~C.,  {Benson} P.~J.,  1983, \mn@doi [\apj] {10.1086/160780}, \href
  {https://ui.adsabs.harvard.edu/abs/1983ApJ...266..309M} {266, 309}

\bibitem[\protect\citeauthoryear{{Nam}, {Federrath}  \& {Krumholz}}{{Nam}
  et~al.}{2021}]{NamFederrathKrumholz2021}
{Nam} D.~G.,  {Federrath} C.,   {Krumholz} M.~R.,  2021, \mn@doi [\mnras]
  {10.1093/mnras/stab505}, \href
  {https://ui.adsabs.harvard.edu/abs/2021MNRAS.503.1138N} {503, 1138}

\bibitem[\protect\citeauthoryear{{Ossenkopf} \& {Mac Low}}{{Ossenkopf} \& {Mac
  Low}}{2002}]{OssenkopfMacLow2002}
{Ossenkopf} V.,  {Mac Low} M.-M.,  2002, \mn@doi [\aap]
  {10.1051/0004-6361:20020629}, \href
  {http://cdsads.u-strasbg.fr/abs/2002A%26A...390..307O} {390, 307}

\bibitem[\protect\citeauthoryear{{Padoan} \& {Nordlund}}{{Padoan} \&
  {Nordlund}}{2002}]{PadoanNordlund2002}
{Padoan} P.,  {Nordlund} {\AA}.,  2002, \mn@doi [\apj] {10.1086/341790}, 576,
  870

\bibitem[\protect\citeauthoryear{{Padoan} \& {Nordlund}}{{Padoan} \&
  {Nordlund}}{2011}]{PadoanNordlund2011}
{Padoan} P.,  {Nordlund} {\AA}.,  2011, \mn@doi [\apj]
  {10.1088/0004-637X/730/1/40}, 730, 40

\bibitem[\protect\citeauthoryear{{Padoan}, {Federrath}, {Chabrier}, {Evans},
  {Johnstone}, {J{\o}rgensen}, {McKee}  \& {Nordlund}}{{Padoan}
  et~al.}{2014}]{PadoanEtAl2014}
{Padoan} P.,  {Federrath} C.,  {Chabrier} G.,  {Evans} II N.~J.,  {Johnstone}
  D.,  {J{\o}rgensen} J.~K.,  {McKee} C.~F.,   {Nordlund} {\AA}.,  2014, in
  {Beuther} H.,  {Klessen} R.~S.,  {Dullemond} C.~P.,   {Henning} T.,  eds,
  Protostars and Planets VI. University of Arizona Press, pp 77--100
  (\mn@eprint {arXiv} {1312.5365}),
  \mn@doi{10.2458/azu_uapress_9780816531240-ch004}

\bibitem[\protect\citeauthoryear{{Ricker}}{{Ricker}}{2008}]{Ricker2008}
{Ricker} P.~M.,  2008, \mn@doi [\apjs] {10.1086/526425}, \href
  {http://cdsads.u-strasbg.fr/abs/2008ApJS..176..293R} {176, 293}

\bibitem[\protect\citeauthoryear{{Roman-Duval}, {Federrath}, {Brunt}, {Heyer},
  {Jackson}  \& {Klessen}}{{Roman-Duval} et~al.}{2011}]{RomanDuvalEtAl2011}
{Roman-Duval} J.,  {Federrath} C.,  {Brunt} C.,  {Heyer} M.,  {Jackson} J.,
  {Klessen} R.~S.,  2011, \mn@doi [\apj] {10.1088/0004-637X/740/2/120}, \href
  {http://adsabs.harvard.edu/abs/2011ApJ...740..120R} {740, 120}

\bibitem[\protect\citeauthoryear{{Roy} \& {Joncas}}{{Roy} \&
  {Joncas}}{1985}]{RoyJoncas1985}
{Roy} J.~R.,  {Joncas} G.,  1985, \mn@doi [\apj] {10.1086/162772}, \href
  {https://ui.adsabs.harvard.edu/abs/1985ApJ...288..142R} {288, 142}

\bibitem[\protect\citeauthoryear{{Schneider} et~al.,}{{Schneider}
  et~al.}{2013}]{SchneiderEtAl2013}
{Schneider} N.,  et~al., 2013, \mn@doi [\apjl] {10.1088/2041-8205/766/2/L17},
  \href {http://adsabs.harvard.edu/abs/2013ApJ...766L..17S} {766, L17}

\bibitem[\protect\citeauthoryear{{Sharda}, {Federrath}, {da Cunha}, {Swinbank}
  \& {Dye}}{{Sharda} et~al.}{2018}]{ShardaEtAl2018}
{Sharda} P.,  {Federrath} C.,  {da Cunha} E.,  {Swinbank} A.~M.,   {Dye} S.,
  2018, \mn@doi [\mnras] {10.1093/mnras/sty886}, \href
  {http://adsabs.harvard.edu/abs/2018MNRAS.477.4380S} {477, 4380}

\bibitem[\protect\citeauthoryear{{Sharda} et~al.,}{{Sharda}
  et~al.}{2019}]{ShardaEtAl2019}
{Sharda} P.,  et~al., 2019, \mn@doi [\mnras] {10.1093/mnras/stz1543}, \href
  {https://ui.adsabs.harvard.edu/abs/2019MNRAS.487.4305S} {487, 4305}

\bibitem[\protect\citeauthoryear{{Sharda}, {Federrath}, {Krumholz}  \&
  {Schleicher}}{{Sharda} et~al.}{2021}]{ShardaEtAl2021}
{Sharda} P.,  {Federrath} C.,  {Krumholz} M.~R.,   {Schleicher} D. R.~G.,
  2021, \mn@doi [\mnras] {10.1093/mnras/stab531}, \href
  {https://ui.adsabs.harvard.edu/abs/2021MNRAS.503.2014S} {503, 2014}

\bibitem[\protect\citeauthoryear{{Solomon}, {Rivolo}, {Barrett}  \&
  {Yahil}}{{Solomon} et~al.}{1987}]{SolomonEtAl1987}
{Solomon} P.~M.,  {Rivolo} A.~R.,  {Barrett} J.,   {Yahil} A.,  1987, \mn@doi
  [\apj] {10.1086/165493}, \href
  {http://cdsads.u-strasbg.fr/abs/1987ApJ...319..730S} {319, 730}

\bibitem[\protect\citeauthoryear{{Stone}, {Ostriker}  \& {Gammie}}{{Stone}
  et~al.}{1998}]{StoneOstrikerGammie1998}
{Stone} J.~M.,  {Ostriker} E.~C.,   {Gammie} C.~F.,  1998, \mn@doi [\apjl]
  {10.1086/311718}, \href {http://cdsads.u-strasbg.fr/abs/1998ApJ...508L..99S}
  {508, L99}

\bibitem[\protect\citeauthoryear{{Sur}, {Schlei\-cher}, {Banerjee}, {Federrath}
   \& {Klessen}}{{Sur} et~al.}{2010}]{SurEtAl2010}
{Sur} S.,  {Schlei\-cher} D.~R.~G.,  {Banerjee} R.,  {Federrath} C.,
  {Klessen} R.~S.,  2010, \mn@doi [\apjl] {10.1088/2041-8205/721/2/L134}, \href
  {http://adsabs.harvard.edu/abs/2010ApJ...721L.134S} {721, L134}

\bibitem[\protect\citeauthoryear{{Turk}, {Smith}, {Oishi}, {Skory}, {Skillman},
  {Abel}  \& {Norman}}{{Turk} et~al.}{2011}]{TurkEtAl2011}
{Turk} M.~J.,  {Smith} B.~D.,  {Oishi} J.~S.,  {Skory} S.,  {Skillman} S.~W.,
  {Abel} T.,   {Norman} M.~L.,  2011, \mn@doi [\apjs]
  {10.1088/0067-0049/192/1/9}, \href
  {http://adsabs.harvard.edu/abs/2011ApJS..192....9T} {192, 9}

\bibitem[\protect\citeauthoryear{{Varidel}, {Pracy}, {Croom}, {Owers}  \&
  {Sadler}}{{Varidel} et~al.}{2016}]{VaridelEtAl2016}
{Varidel} M.,  {Pracy} M.,  {Croom} S.,  {Owers} M.~S.,   {Sadler} E.,  2016,
  \mn@doi [\pasa] {10.1017/pasa.2016.3}, \href
  {https://ui.adsabs.harvard.edu/abs/2016PASA...33....6V} {33, e006}

\bibitem[\protect\citeauthoryear{{W{\"u}nsch}, {Walch}, {Dinnbier}  \&
  {Whitworth}}{{W{\"u}nsch} et~al.}{2018}]{WuenschEtAl2018}
{W{\"u}nsch} R.,  {Walch} S.,  {Dinnbier} F.,   {Whitworth} A.,  2018, \mn@doi
  [\mnras] {10.1093/mnras/sty015}, \href
  {https://ui.adsabs.harvard.edu/abs/2018MNRAS.475.3393W} {475, 3393}

\bibitem[\protect\citeauthoryear{{Zhou} et~al.,}{{Zhou}
  et~al.}{2017}]{ZhouEtAl2017}
{Zhou} L.,  et~al., 2017, \mn@doi [\mnras] {10.1093/mnras/stx1504}, \href
  {http://adsabs.harvard.edu/abs/2017MNRAS.470.4573Z} {470, 4573}

\makeatother
\end{thebibliography}




\appendix

\section{Moment maps for LOS along the rotation axis of the cloud}

\begin{figure*}
\centering
\includegraphics[width=\textwidth]{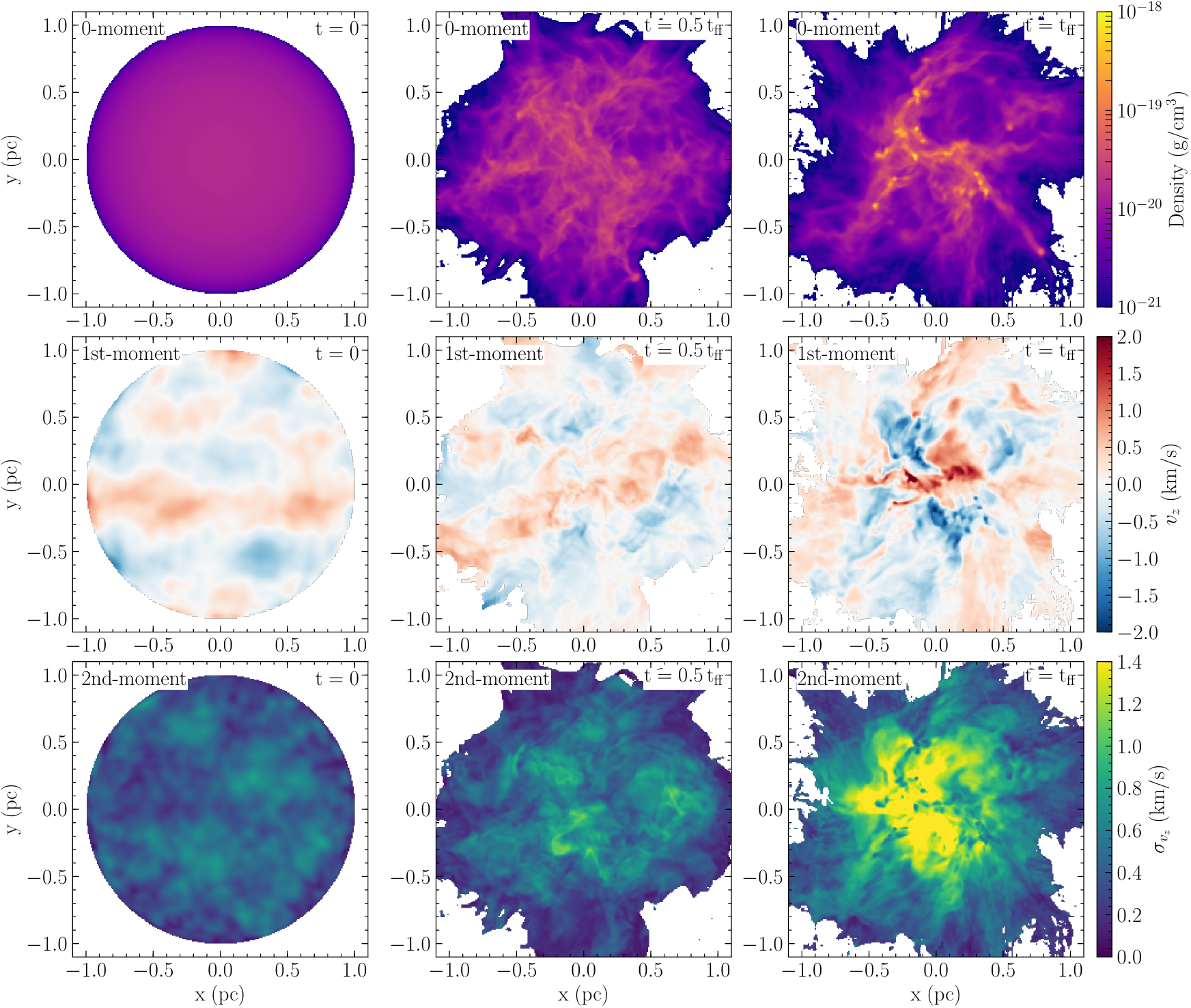}
\caption{Same as Fig.~\ref{fig: moment maps}, but for the LOS along the rotation axis ($z$ axis). The main difference is in the 1st-moment map, which does not show a gradient as we are here looking along the rotation axis.}
\label{fig: moment maps z-axis}
\end{figure*}

Figure~\ref{fig: moment maps z-axis} shows the same as Fig.~\ref{fig: moment maps}, but for the LOS along the $z$ direction, i.e., along the rotation axis of the cloud. The cloud's shape remains fairly symmetrical throughout the evolution of the cloud in this projection, unlike when the cloud is viewed along the $x$ or $y$ axes, in which case a flattening of the cloud along the $z$-axis due to rotation is visible (Fig.~\ref{fig: moment maps}). The most notable difference in these moment maps compared to the moment maps viewed along the $y$-axis (Fig.~\ref{fig: moment maps}) is the lack of a visible gradient in the 1st-moment map, which we expect because the $z$-axis is the rotation axis.

\section{Gradient subtraction in case of LOS along the rotation axis \label{sec: gradients and first moments z axis}}

\begin{figure*}
\centering
\includegraphics[width=\textwidth]{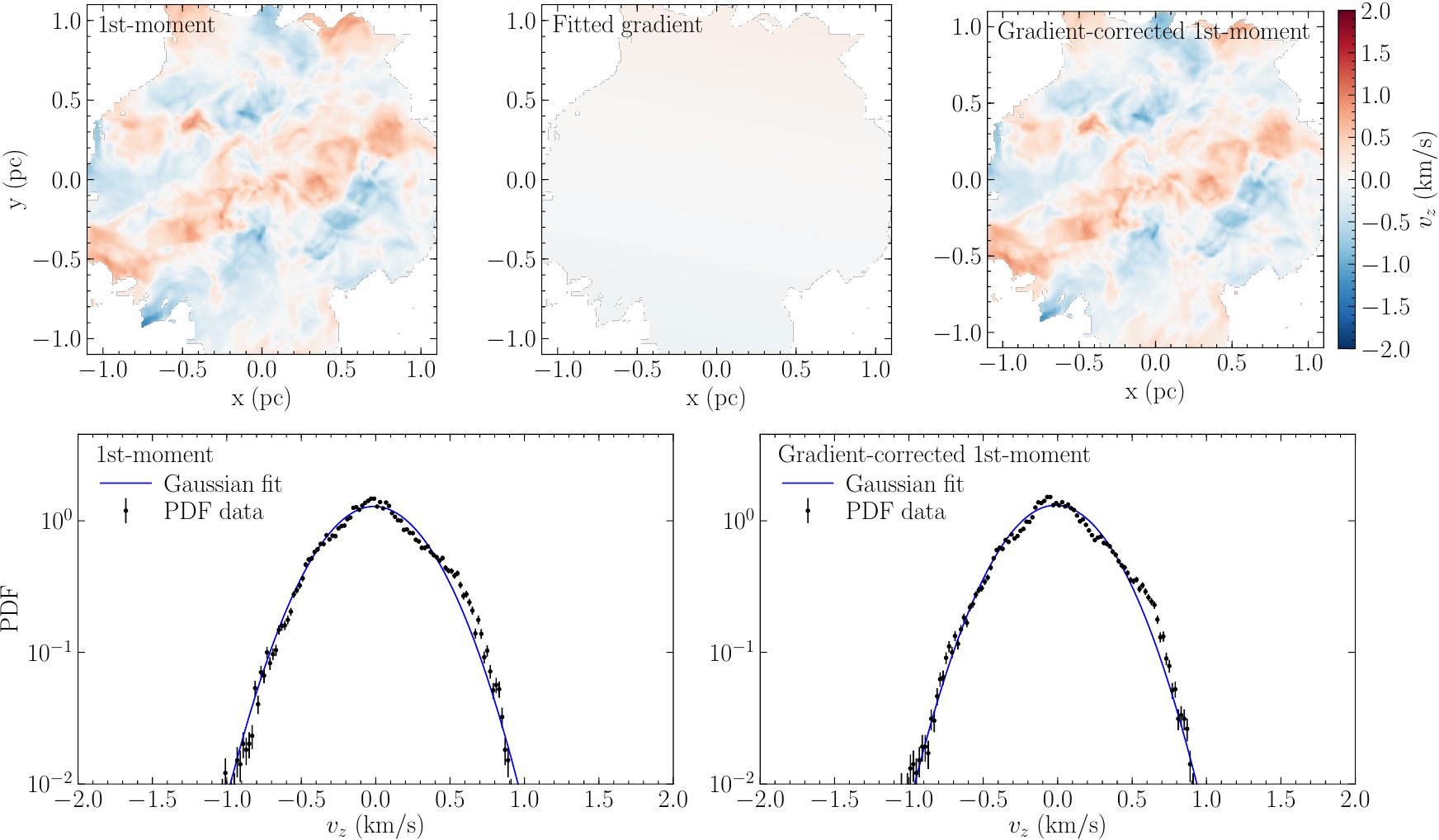}
\caption{Same as Fig.~\ref{fig: gradients and first moments y axis}, but viewed along the rotation axis ($z$). The fitted gradient in this case is nearly zero, and hence the gradient-subtracted 1st moment (top right) is practically identical to the original 1st moment (top left). The same applies to the velocity PDFs (bottom panels), which are the same before and after subtraction of the gradient, i.e., there is no contribution of rotation when the LOS is oriented along the rotation axis.}
\label{fig: gradients and first moments z axis}
\end{figure*}

 Figure~\ref{fig: gradients and first moments z axis} illustrates the process of fitting a gradient to the 1st-moment map of a cloud viewed along the rotation axis, and subtracting this fitted gradient. The gradient fit to the 1st-moment map (top middle panel) is negligible in this case, so its subtraction makes little difference to the rotation-corrected 1st-moment map. This is further quantified in the velocity PDFs shown in the bottom panels, which are practically identical before and after gradient subtraction.

\section{Correction factors with a higher cutoff threshold based on the zero-moment map} \label{sec: appendix cutoff 1}

\begin{figure*}
\centering
\includegraphics[width=\textwidth]{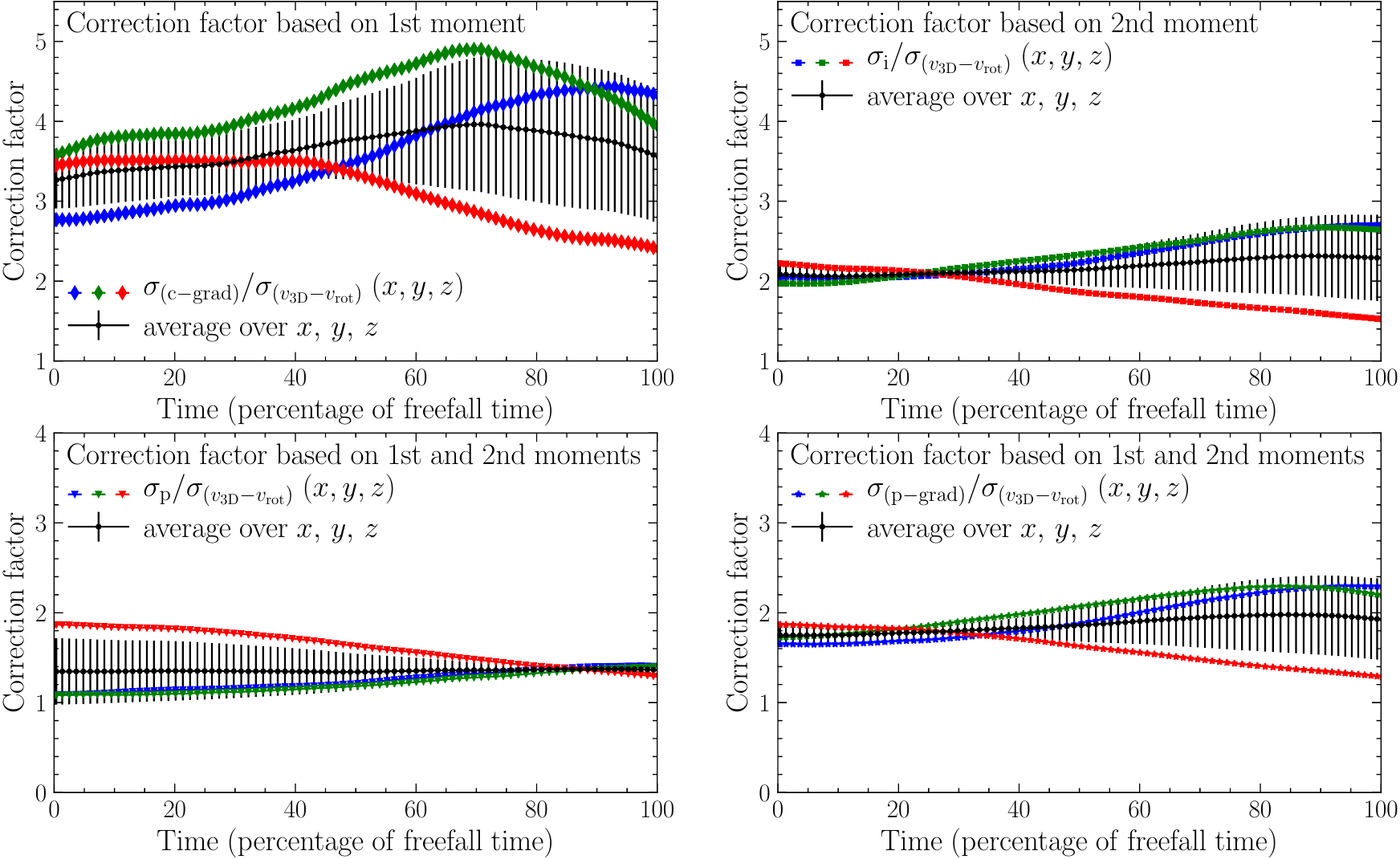}
\caption{Same as Fig.~\ref{fig: correction factors}, but with a column-density cutoff based on the average column density (instead of 10\% of the average column density), i.e., here we exclude cloud material with a column density below the average column density of the cloud.}
\label{fig: correction factors cutoff1}
\end{figure*}

Figure~\ref{fig: correction factors cutoff1} shows the same as Fig.~\ref{fig: correction factors}, but for a cutoff of every pixel with a column density less than the average column density of the cloud being excluded. We see that using a threshold of 1 instead of 0.1 for the cutoff does not significantly affect the time average of the correction factors, but it does increase the uncertainty in these correction factors. Comparing this to the corresponding version of this figure with the lower (0.1 average column density) cutoff (Figure~\ref{fig: correction factors}), we see that the differences in correction factors between the different LOS axes is greater with the higher cutoff in Figure~\ref{fig: correction factors cutoff1}, more so for the standard deviation of the gradient-corrected 1st-moment map than the spatial mean of the 2nd-moment map. The spatial mean of the 2nd-moment map appears to be more stable in time for the higher cutoff, and so is the gradient-corrected parent dispersion.

\section{Effects of telescope beam resolution on second moment} \label{sec: second moment fwhm dependence}
\begin{figure*}
    \centering
    \includegraphics[width=0.5\linewidth]{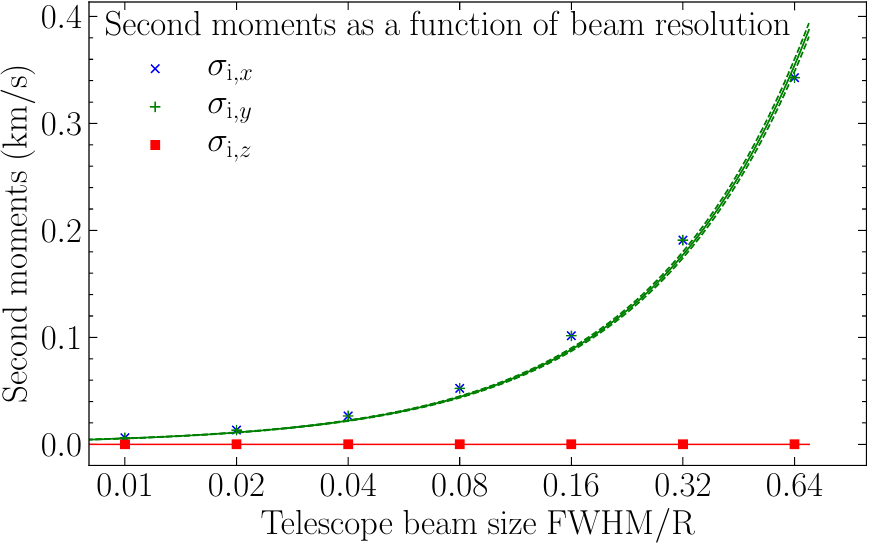}
    \caption{Spatial mean of the 2nd-moment map as a function of telescope beam size as measured along the $x$ (blue), $y$ (green) and $z$ (red) axes. The 2nd moments are measured for a cloud with an angular frequency of $\Omega=5.124 \times 10^{-14}\,\s^{-1}$, and no turbulence, to demonstrate the effect of beam smearing. Note that the $x$ and $y$ axes yield identical results here due to the symmetry of the idealised cloud. A linear function passing through the origin is fitted to the 2nd moments measured along the $x$ and $y$ axes, giving $\sigtwoKD{x,y} = f/R\times (0.55 \pm 0.01)\,\mathrm{km/s}$, where $f$ is the telescope beam FWHM and $R$ is the cloud radius. The 2nd moment increases with increasing telescope beam size when viewed along any axis other than the rotation axis.}
    \label{fig: second moment vs fwhm}
\end{figure*}

Considering an ideal cloud without turbulence and solid-body rotation only, the line of sight velocities along an infinitesimally small beam will all be the same. This means that the 2nd moment will be zero for each pixel in the plane of the sky. As the telescope beam size increases, the velocities of the gas measured in a given beam will include more and more contributions from velocities at different plane-of-sky positions. The mixing of these different LOS velocities inside the growing beam size induces a beam-dependent dispersion. Hence, the spatial mean of the 2nd moment of a rotating cloud increases with increasing telescope beam size. This effect is quantified in Fig.~\ref{fig: second moment vs fwhm}, which shows that as the cloud is observed perpendicular to its rotation axis (LOS along $x$ or $y$), the spatial mean of the 2nd-moment map increases as the telescope beam size increases. If viewed along the rotation axis (LOS along $z$) there is no effect of rotation and the 2nd moment remains identically zero, as expected.

\section{Fit parameters for correction factors as a function of telescope beam size} \label{sec: lines of best fit}

\begin{table*}
    \centering
    \caption{Correction factors as a function of telescope beam FWHM for different LOS axes and simulations with different levels of rotation. Parabolic functions are fitted to the correction factors as a function of telescope beam FWHM ($f$), as displayed in Figures~\ref{fig: beam resolution study} and \ref{fig: effects of rotation}. The fit parameters are given here, where the line of best fit is defined as $C_\mathrm{statistic}^\mathrm{LOS}(f/R) = p_2 \left( f / R \right)^2 + p_1 \left(f/R\right) + p_0$. In the LOS axis column and in the $E_\mathrm{rot} / E_\mathrm{turb}$ column, `any' indicates functions that are fit to data from all three LOS axes combined, and/or all three simulations combined, respectively. Note that a linear function instead of a quadratic function is fitted to the data in cases where a linear fit describes the data well; in such cases, $p_2 = 0$.}
    \def\arraystretch{1.4}
    \begin{tabular}{lllcccc}
    \hline
        Correction factor & Statistic & LOS axis & $E_\mathrm{rot} / E_\mathrm{turb}$ & $p_2$ & $p_1$ & $p_0$ \\
    \hline
$C_{\mathrm{(c} - \mathrm{grad)}}^{x}$ for $E_\mathrm{rot}/E_\mathrm{turb} = 1.00$ & $\siggradKDgen$ & $x$ & 1.00 & $5.54 \pm 2.99$ & $\phantom{-}3.30 \pm 1.96$ & $2.89 \pm 0.36$\\
$C_{\mathrm{(c} - \mathrm{grad)}}^{y}$ for $E_\mathrm{rot}/E_\mathrm{turb} = 1.00$ & $\siggradKDgen$ & $y$ & 1.00 & $3.58 \pm 3.25$ & $\phantom{-}4.24 \pm 2.14$ & $3.12 \pm 0.39$\\
$C_{\mathrm{(c} - \mathrm{grad)}}^{z}$ for $E_\mathrm{rot}/E_\mathrm{turb} = 1.00$ & $\siggradKDgen$ & $z$ & 1.00 & $20.8 \pm 2.1\phantom{0}$ & $\phantom{-}3.20 \pm 1.41$ & $3.26 \pm 0.27$\\
$C_{\mathrm{(c} - \mathrm{grad)}}^\mathrm{any}$ for $E_\mathrm{rot}/E_\mathrm{turb} = 0.00$ & $\siggradKDgen$ & any & 0.00 & $20.2 \pm 10.7$ & $\phantom{-}6.01 \pm 7.05$ & $3.66 \pm 1.30$\\
$C_{\mathrm{(c} - \mathrm{grad)}}^\mathrm{any}$ for $E_\mathrm{rot}/E_\mathrm{turb} = 0.35$ & $\siggradKDgen$ & any & 0.35 & $13.6 \pm 5.7\phantom{0}$ & $\phantom{-}3.45 \pm 3.76$ & $3.22 \pm 0.70$\\
$C_{\mathrm{(c} - \mathrm{grad)}}^\mathrm{any}$ for $E_\mathrm{rot}/E_\mathrm{turb} = 1.00$ & $\siggradKDgen$ & any & 1.00 & $10.1 \pm 4.3\phantom{0}$ & $\phantom{-}3.48 \pm 2.85$ & $3.09 \pm 0.55$\\
$C_\mathrm{i}^{x}$ for $E_\mathrm{rot}/E_\mathrm{turb} = 1.00$ & $\sigtwoKDgen$ & $x$ & 1.00 & $0$ & $-1.21 \pm 0.51$ & $2.40 \pm 0.29$\\
$C_\mathrm{i}^{y}$ for $E_\mathrm{rot}/E_\mathrm{turb} = 1.00$ & $\sigtwoKDgen$ & $y$ & 1.00 & $0$ & $-1.24 \pm 0.46$ & $2.50 \pm 0.27$\\
$C_\mathrm{i}^{z}$ for $E_\mathrm{rot}/E_\mathrm{turb} = 1.00$ & $\sigtwoKDgen$ & $z$ & 1.00 & $0$ & $-0.74 \pm 0.18$ & $2.20 \pm 0.10$\\
$C_\mathrm{i}^\mathrm{any}$ for $E_\mathrm{rot}/E_\mathrm{turb} = 0.00$ & $\sigtwoKDgen$ & any & 0.00 & $0$ & $-0.87 \pm 0.25$ & $2.31 \pm 0.14$\\
$C_\mathrm{i}^\mathrm{any}$ for $E_\mathrm{rot}/E_\mathrm{turb} = 0.35$ & $\sigtwoKDgen$ & any & 0.35 & $0$ & $-0.97 \pm 0.28$ & $2.34 \pm 0.16$\\
$C_\mathrm{i}^\mathrm{any}$ for $E_\mathrm{rot}/E_\mathrm{turb} = 1.00$ & $\sigtwoKDgen$ & any & 1.00 & $0$ & $-1.07 \pm 0.45$ & $2.37 \pm 0.26$\\
$C_{\mathrm{(p} - \mathrm{grad)}}^\mathrm{any}$ for $E_\mathrm{rot}/E_\mathrm{turb} = 1.00$ & $\sigpgradKDgen$ & any & 1.00 & $0$ & $-0.31 \pm 0.25$ & $1.90 \pm 0.14$\\
$C_{\mathrm{(p} - \mathrm{grad)}}^\mathrm{any}$ for any $E_\mathrm{rot}/E_\mathrm{turb}$ & $\sigpgradKDgen$ & any & any & $0$ & $-0.29 \pm 0.26$ & $1.93 \pm 0.15$\\
    \end{tabular}
    \label{tab: lines of best fit}
\end{table*}

The fit parameters for the relationship between the correction factors for the three potential turbulence quantifier statistics and the telescope beam FWHM, as displayed in Fig~\ref{fig: beam resolution study} and Fig~\ref{fig: effects of rotation}. These functions can be used to determine the correction factor necessary to recover the 3D turbulent velocity dispersion from a given statistic with a given telescope beam size. The function we anticipate to be the most useful to observers is the function for $\sigpgradKDgen$ fit to the correction factors for all LOS and all simulations simultaneously ($C^\mathrm{any}_{\mathrm{(p}-\mathrm{grad)}}$ for any $E_\mathrm{rot}/E_\mathrm{turb}$). The other functions may be useful in a situation in which insufficient data is available, for example, if only the second moment data were accessible.

\bsp	
\label{lastpage}
\end{document}